\documentclass[aps, prx, twocolumn, superscriptaddress, nofootinbib,
               floatfix, longbibliography]{revtex4-2}

\usepackage{graphicx}
\usepackage{amsmath}
\usepackage{amssymb}
\usepackage{bm}
\usepackage[colorlinks=true, linkcolor=blue, citecolor=blue,
            urlcolor=blue]{hyperref}

\graphicspath{{figs/}}

\newcommand{\wv}[1]{\left(#1\right)_{w}}
\newcommand{\ket}[1]{|#1\rangle}
\newcommand{\bra}[1]{\langle#1|}
\newcommand{\braket}[2]{\langle#1|#2\rangle}
\newcommand{\gfwd}{\gamma_{\rm fwd}}
\newcommand{\gbwd}{\gamma_{\rm bwd}}
\newcommand{\gA}{\gamma_{A}}
\newcommand{\gS}{\gamma_{S}}
\newcommand{\dtcg}{\Delta t_{\rm cg}}

\begin{document}

\title{Two-state generator extraction: property currents and a two-layer
arrow of time in pre- and post-selected quantum dynamics}

\author{Seiki Saito}
\email{saitos@yz.yamagata-u.ac.jp}
\affiliation{Graduate School of Science and Engineering, Yamagata University,
Yonezawa 992-8510, Japan}

\date{\today}

\begin{abstract}
An observer who conditions on both the past and the future assigns physical
properties to intermediate times that a causally conditioned observer does not.
We show that these time-symmetric assignments are governed by exact symmetry
theorems, and that the resulting structure is directly measurable from
trajectory data. Our tool is two-state generator extended dynamic mode
decomposition (gEDMD): because weak values obey
$dA_{w}/dt = i\langle[H,A]\rangle_{w}$ exactly, the machinery of data-driven
generator extraction (dictionary regression, finite-difference estimators, and an exact-derivative baseline) applies unchanged to complex
weak-value trajectories. Two consequences follow. First, a single reflection
involution on the pre-/post-selected ensemble forces every window-fitted
friction to split uniquely as
$\gamma_{\rm fwd} = \gamma_{A} + \gamma_{S}$, with $\gamma_{A}$ antisymmetric
about the midpoint, carrying the boundary-condition physics of the coherent
modes, and $\gamma_{S}$ symmetric and generated by the differencing scheme;
the two components are
separable from the same data as $(\gamma_{\rm fwd} \pm
\gamma_{\rm bwd})/2$. At a fixed inference resolution the ``arrow of time''
therefore has two layers: the
coherent-mode arrow reverses at the midpoint, whereas the fluctuation-level
arrow does not, because $\gamma_{S}$ dominates $\gamma_{A}$ there at every size
and conditioning class we test. The asymmetry between the layers is one of
degree, the scheme component being $34$ times larger at the fluctuation layer
than at the mode layer, and with the exact derivative both layers reverse; the
immunity of the second layer is therefore a property of the inference and not
of the ensemble. Second, in a lattice interferometer whose pre- and post-selections
are specified only at the input and output ports, the canonical quantum
Cheshire-cat structure emerges without being imposed, with the particle and its
polarization obeying separate continuity equations; a local field in the arm
carrying the polarization rotates only the polarization's phase, at exactly
twice the field strength, and appears in the extracted generator as a rigid
imaginary spectral shift, while the particle's weak density stays invariant to
machine precision. We verify the sample-level involution identity and the
two-layer structure from $2^{8}$ to $2^{20}$ Hilbert-space dimensions, where
the no-flip inequality holds with $|\gamma_{A}|/\gamma_{S} = 0.09$--$0.27$
across five post-selection classes ($0.135$ at $2^{8}$), and where
self-averaging renders the result independent of the post-selection class
among classes sharing the same boundary modulation, which removes the
$|\braket{\varphi}{\psi}| \sim 2^{-N/2}$ obstruction to scaling for $N$ qubits.
\end{abstract}

\maketitle

\section{Introduction}
\label{sec:intro}

Whether an isolated quantum system exhibits friction is not a property of the
system alone. In a companion study~\cite{Saito2026sister} we showed that the
macroscopic friction extracted from the unitary dynamics of a chaotic spin
chain is generated by the observer's temporal resolution and by the causal
direction of the coarse-grained inference: a time-symmetric estimator returns
zero net dissipation at any coarse-graining scale, while a forward-in-time
estimator returns a strictly positive friction. Friction, in that sense, is a
statement about the description rather than about the state.

That conclusion was reached for a causally conditioned observer, who knows the
past and predicts forward. Quantum mechanics, however, admits a strictly larger
class of descriptions. In the two-state-vector
formalism~\cite{ABL1964, AAV1988}, an observer who conditions on a
post-selection at a later time $T$ assigns to an intermediate time $t$ the
weak value
\begin{equation}
  A_{w}(t) = \frac{\bra{\varphi(t)} A \ket{\psi(t)}}
                  {\braket{\varphi(t)}{\psi(t)}},
  \label{eq:wv-def}
\end{equation}
which is in general complex and which can lie far outside the spectrum of $A$.
We write $A_{w}$, $\wv{A}$ and $\langle A\rangle_{w}$ interchangeably for this
weak value, while an unsubscripted $\langle A \rangle = \bra{\psi}A\ket{\psi}$
denotes the ordinary (causally conditioned) expectation value.
Such time-symmetric, or \emph{smoothed}, descriptions are operationally
meaningful: they are the objects estimated by weak measurements, and within the
time-symmetric theory of quantum smoothing~\cite{Tsang2009} the least-squares
smoothed estimator of an observable equals
$\mathrm{Re}\,A_{w}$~\cite{Gammelmark2013, Chantasri2021}. They also
support assignments with no causal counterpart. The best-known example is the
quantum Cheshire cat~\cite{Aharonov2013cheshire, Denkmayr2014}, in which a
particle and one of its properties are assigned to different arms of an
interferometer, an effect whose dynamical extension, a current of the property unaccompanied by a current of the particle, was proposed by
Aharonov, Cohen and Popescu~\cite{Aharonov2015current, Aharonov2021dynamical}
and whose interpretation remains
contested~\cite{Correa2015, Duprey2018, Hance2024detectable,
Hance2023contextuality}. Related separations have been demonstrated
experimentally, in an exchange of properties between two such systems~\cite{Liu2020grins}, and proposed for a particle's mass and
momentum~\cite{Waegell2024}.

This raises the question that the present work addresses. If friction and the
arrow of time are functions of the observer's description, what happens when
the description is extended from the causal to the time-symmetric class? Which
of the assignments an observer makes at intermediate times can be altered by
conditioning on the future, and which cannot?

Answering this requires a tool that extracts effective dynamics from
conditioned trajectories without presupposing their form. We provide one. The
key observation is that the weak value~\eqref{eq:wv-def} is linear in $A$ and,
when $\ket{\psi}$ and $\ket{\varphi}$ propagate under the same Hamiltonian,
satisfies
\begin{equation}
  \frac{d A_{w}}{dt} = i \left\langle [H, A] \right\rangle_{w}
  \label{eq:exact-deriv}
\end{equation}
exactly, with $\braket{\varphi(t)}{\psi(t)}$ constant in time. Equation
\eqref{eq:exact-deriv} has the same structure as the Heisenberg equation for
ordinary expectation values, and therefore the entire apparatus of data-driven
generator extraction, namely extended dynamic mode decomposition in its
generator form (gEDMD)~\cite{Williams2015, Klus2018, Klus2020}, combined with a Markovian
truncation of the Mori--Zwanzig hierarchy~\cite{Mori1965, Zwanzig1973}, transfers unchanged to complex
weak-value time series. We refer to this as \emph{two-state gEDMD}. Its
key feature is not the regression itself but the fact that
Eq.~\eqref{eq:exact-deriv} supplies an \emph{exact-derivative baseline}: a
reference against which any finite-difference estimator can be compared, so
that ``physics'' and ``artifact of the estimator'' can be told apart. In the
generic data-driven setting no such reference exists; here the two-state
structure supplies one even though the trajectories are conditioned.

With this tool we establish two results. The first concerns \emph{where a
property is}. We construct a lattice interferometer in which the pre- and
post-selections are specified only at the input and output ports, and show that
the canonical Cheshire-cat structure then emerges without being imposed by
hand, at the level of conservation laws: the particle and its polarization each
carry their own continuity equation with their own current. We then make the
assignment operational by perturbing it. A field placed in the arm to which the
polarization is assigned rotates only the polarization's phase, at exactly twice the field strength (a factor traceable to the two-state structure), while
the particle's weak density remains invariant to machine precision; the effect
is absent from the causal description and from the post-selection probability
alike. Extracted as a generator, the field appears as a rigid imaginary shift
of the property sector's spectrum, with the particle sector untouched. In that
same generator we find the dichotomy of Ref.~\cite{Saito2026sister} again: the
precession rate is estimator-independent, while an apparent friction is
introduced by the forward difference.

The second result promotes that dichotomy to a theorem. For a broad class of
pre-/post-selected ensembles there is a reflection involution,
$(\ket{\psi_{0}}, \ket{\varphi_{T}}) \mapsto (\ket{\varphi_{T}^{*}},
\ket{\psi_{0}^{*}})$, which preserves the $|\braket{\varphi}{\psi}|^{2}$
weighted ensemble exactly (Appendix~\ref{app:proof}). Equivariance of the
window-fitting estimators under this involution forces
$\gfwd(t) = -\gbwd(T-t)$ and hence the unique decomposition
\begin{equation}
  \gfwd = \underbrace{\gA}_{\text{antisymmetric}}
        + \underbrace{\gS}_{\text{symmetric}},
  \label{eq:decomp-intro}
\end{equation}
in which $\gA$ agrees with the exact-derivative friction and carries the
boundary-condition arrow of the coherent modes, while $\gS$ is generated by
the differencing scheme and cannot reverse across the midpoint; at the fluctuation layer the
small residual $\gA$ is dictionary-dependent, and only the inequality
$\gS > |\gA|$ enters our claims. The ``two-layer arrow of time'' is then a corollary
rather than a separate phenomenon: at the level of coherent modes $\gA$
dominates and the arrow reverses at the midpoint of the conditioning window,
whereas at the level of the fluctuation layer $\gS$ dominates and no reversal
is seen. That second statement is made at a fixed differencing resolution:
$\gS$ is what the estimator injects and vanishes with the coarse-graining
interval, so with the exact derivative the fluctuation layer reverses as well,
and the two layers are distinguished by how much scheme bias each receives
rather than by the presence or absence of a reversing component
(Sec.~\ref{sec:resII-twolayer}). Because $\gA$ and $\gS$ are obtained from the same data as
$(\gfwd \pm \gbwd)/2$, the physical and the inferential arrow are
\emph{operationally separable}.

Finally, we ask whether any of this survives at scale. We verify the involution
and the two-layer structure in a chaotic XXZ chain from $N=8$ to $N=20$
qubits, i.e.\ from $2^{8}$ to $2^{20}$ Hilbert-space dimensions. The
sample-level reflection identity holds at the $10^{-14}$--$10^{-13}$ level
across the modes and a $77$-dimensional hydrodynamic dictionary, confirming
that the theorem is an algebraic consequence and not a small-system
coincidence; the no-flip inequality of the fluctuation layer holds with a
comparable margin at both sizes, $|\gA|/\gS = 0.135$ at $N=8$ and
$0.09$--$0.27$ across five post-selection classes at $N=20$. At the larger size we find in addition
that self-averaging renders the result insensitive to the post-selection
class: the four bright classes collapse onto one another at the $10^{-3}$
level or below, the independent
class agrees within its statistical precision, and for independently drawn
post-selections this follows from the structure of the weighted average, which
depends on the conditioning only through its second moment. This has a practical consequence: the
brightest available class may be used without changing the physics, which
removes the $|\braket{\varphi}{\psi}| \sim 2^{-N/2}$ obstruction that otherwise
makes independent post-selection unusable in large systems.

The paper is organized as follows. Section~\ref{sec:method} develops two-state
gEDMD. Section~\ref{sec:resI} presents the interferometer and the response
experiment. Section~\ref{sec:resII} establishes the reflection involution, the
decomposition~\eqref{eq:decomp-intro} and the two layers.
Section~\ref{sec:resIII} scales the analysis to $2^{20}$ dimensions.
Section~\ref{sec:discussion} discusses the relation to the literature on
smoothed quantum states and to the Cheshire-cat debate, and
Sec.~\ref{sec:conclusion} concludes.

\section{Two-state gEDMD}
\label{sec:method}

\subsection{Exact-derivative identity}
\label{sec:method-identity}

Let $\ket{\psi(t)}$ and $\ket{\varphi(t)}$ both evolve under the same,
possibly time-dependent, Hamiltonian $H(t)$, the former forward from a
pre-selection $\ket{\psi_{0}}$ at $t=0$ and the latter backward from a
post-selection seed $\ket{\varphi_{T}}$ at $t=T$. Unitarity of the shared
propagator makes the overlap
$g \equiv \braket{\varphi(t)}{\psi(t)}$ independent of $t$, and for any
time-independent observable $A$ the weak value~\eqref{eq:wv-def} obeys
Eq.~\eqref{eq:exact-deriv} identically. Two remarks are in order. First, the
identity holds for time-dependent $H$, which is what allows us to treat
interferometers with active elements in Sec.~\ref{sec:resI}. Second, taking
$\ket{\varphi_{T}} = \ket{\psi(T)}$ makes the post-selection trivial and
reduces $A_{w}$ to the ordinary expectation value; we use this as a built-in
consistency check throughout.

\subsection{Constructive currents and conservation laws}
\label{sec:method-currents}

Effective transport statements require currents. That weak values support a
continuity equation is not new: the weak value of the velocity is the Bohmian
velocity, and the associated flow is single-valued and
operational~\cite{Wiseman2007}, a construction that underlies the weak-value
reconstruction of average photon
trajectories~\cite{Kocsis2011}. What we add is to treat the current of a
\emph{property} on the same footing as the current of the particle, so that the
two obey a pair of conservation laws, and to define both constructively, since sign conventions for currents are a standard source of error. For site projectors
$P_{i}$ (written $n_{i}$ in the interferometer sections) and any internal
operator $S$,
\begin{equation}
  J_{j \to i} \equiv i\left(P_{j} H P_{i} - P_{i} H P_{j}\right) S ,
  \label{eq:current-def}
\end{equation}
which satisfies $\sum_{j} J_{j\to i} = i[H, P_{i}]S$ identically. Writing
$H_{\rm nc}$ for the part of the Hamiltonian that fails to commute with $S$,
and assuming $[H_{\rm nc}, P_{i}] = 0$, which holds because the perturbations considered below are diagonal in position, the continuity equation
\begin{equation}
  \frac{d}{dt}\wv{P_{i} S} = \sum_{j} \wv{J_{j \to i}}
    + \wv{i[H_{\rm nc}, P_{i}S]}
  \label{eq:continuity}
\end{equation}
becomes an operator identity, with the last term, a genuine source equal to $\wv{iP_{i}[H_{\rm nc}, S]}$ under the stated assumption, present only when
$H_{\rm nc} \neq 0$. Equation~\eqref{eq:continuity} is what elevates the Cheshire cat from a
statement about a single instant to a statement about conserved property
currents, and the source term is precisely what carries the response of
Sec.~\ref{sec:resI-response}.

\subsection{Generator regression and differencing schemes}
\label{sec:method-gedmd}

Given a dictionary $\bm{X}(t)$ of weak-value observables we regress a reduced
generator $L$ from $\dot{\bm{X}} \approx L \bm{X}$ by least squares,
minimizing $\sum \| \dot{\bm{X}} - L\bm{X} \|^{2}$ over the data in a window.
This is the same objective as in the causal
case~\cite{Williams2015, Klus2020, Saito2026sister}, where it is equivalent to
a Galerkin projection onto the span of the dictionary. We solve it in the basis
of the leading singular vectors of the (weighted) snapshot matrix, discarding
singular values below a relative threshold, $10^{-9}$ for the interferometer of
Sec.~\ref{sec:resI} and $10^{-6}$ for the many-body dictionaries of
Secs.~\ref{sec:resII}--\ref{sec:resIII}; this truncated singular value
decomposition is the regularization of Ref.~\cite{Williams2015}, needed because
a dictionary evaluated on a finite window is generally close to degenerate. At
the two-element mode layer of Sec.~\ref{sec:resII-setting} the truncation never
activates. The derivative
$\dot{\bm{X}}$ may be supplied in three ways: exactly, from
Eq.~\eqref{eq:exact-deriv}; by the causal forward difference
$[\bm{X}(t+\dtcg) - \bm{X}(t)]/\dtcg$; or by the backward difference. Comparing
the three is the central diagnostic of this work, and the availability of the
exact derivative is what makes the comparison conclusive.

\subsection{Ensembles and \texorpdfstring{$|g|^{2}$}{|g|^2} weighting}
\label{sec:method-weighting}

Throughout this paper ``boundary'' refers to the two ends of the time interval,
$t=0$ and $t=T$, at which the pre- and post-selection are imposed. The
conditioned problem is a boundary-value problem in time, and each boundary
condition is a state of the whole register, not a condition on any subset of
sites; the spatial boundaries of the lattice, which are open, play no role in
what follows and are mentioned only where the Hamiltonian is specified. The two
settings of this paper differ in how these temporal boundary conditions are
supplied, and therefore in whether an ensemble is needed at all. In the
interferometer of Sec.~\ref{sec:resI} the pre- and post-selection are two
specified states, one wave packet at each port, so a single deterministic
trajectory carries the entire result and no averaging occurs anywhere in
Sec.~\ref{sec:resI}: every number quoted there is a property of that one
conditioned trajectory. In the chain of
Secs.~\ref{sec:resII}--\ref{sec:resIII} there is no distinguished pair of
boundary states; what is specified is a \emph{measure} from which they are
drawn, so the objects of interest are ensemble averages and the results are
statistical.

For those many-body ensembles each sample is generated from a pair of random
seed vectors $v$ and $w$, one for the initial time and one for the final time,
each a vector in the full $D$-dimensional space,
\begin{equation}
  \ket{\psi_{0}} = \mathcal{N}[e^{\varepsilon \Lambda} v] , \qquad
  \ket{\varphi_{T}} = \mathcal{N}[e^{\varepsilon \Lambda} w] ,
  \label{eq:seeds}
\end{equation}
where $\mathcal{N}[\cdot]$ denotes normalization to unit norm, $\Lambda$ is a
real diagonal long-wavelength modulation, and $\varepsilon$ sets the
conditioning strength. The seeds $v$ and $w$ are drawn independently of each
other and of all other samples from a conjugation-invariant complex measure,
meaning one under which $v$ and $v^{*}$ are equally likely; concretely we take
the real and imaginary parts of each component to be independent and uniform on
$[-1/2, 1/2]$. Conjugation invariance is what the involution of
Sec.~\ref{sec:resII-involution} requires, and $e^{\varepsilon \Lambda}$ is real
and diagonal, so it preserves that property.

Two properties of Eq.~\eqref{eq:seeds} should be emphasized, because they
recur throughout. First, $\ket{\varphi_{T}}$ is a freshly drawn state at
$t = T$, not the forward-evolved $\ket{\psi(T)}$ nor anything computed from
it: the post-selection carries no information about the pre-selection, which
is exactly the hypothesis the theorem of Sec.~\ref{sec:resII} needs. Second,
the price of that independence is brightness. Two independently drawn states
in $D$ dimensions overlap as
$|\braket{\varphi_{T}}{\psi(T)}| \sim D^{-1/2} = 2^{-N/2}$, so the ensemble is
dim and the $|g|^{2}$ weights concentrate on a few samples, which is the
obstruction that Sec.~\ref{sec:resIII-selfaverage} removes. Equation
\eqref{eq:seeds} therefore defines what we call the independent class; the
other post-selection classes used in Sec.~\ref{sec:resIII} do build
$\ket{\varphi_{T}}$ from $\ket{\psi(T)}$, which makes them bright but breaks
the involution, and they are introduced where they are used. Averages over the $M$ samples of an ensemble must be weighted by
$|g|^{2}$, the post-selection probability,
\begin{equation}
  \overline{A_{w}}
  = \frac{\sum_{s} \braket{\psi_{s}}{\varphi_{s}}
          \bra{\varphi_{s}} A \ket{\psi_{s}}}
         {\sum_{s} |\braket{\varphi_{s}}{\psi_{s}}|^{2}} ,
  \label{eq:selfaverage}
\end{equation}
the overline denoting this average throughout; unweighted averages are
dominated by
samples that would rarely be post-selected. Samples with $|g| < 10^{-2}$ are
rejected at $N=8$ (no rejection is needed at $N=20$, where the weighting
itself suppresses dim samples), a criterion that depends on $|g|$ alone and
therefore respects the involution of Sec.~\ref{sec:resII-involution}.

Two consequences of working with a finite $M$ should be kept in view. First,
the symmetry statements of Sec.~\ref{sec:resII} hold exactly for the measure
the samples are drawn from, and only up to statistical error for any finite
ensemble, which is why we also use
reflection-\emph{paired} sampling, drawing samples in mirror pairs so that the
involution is realized within the ensemble itself and the predicted symmetries
become machine-precision identities rather than statistical statements. Second,
the uncertainty quoted for a half-window average is a bootstrap over samples:
we resample the ensemble with replacement $200$ times, recompute the weighted
mean and refit, and report the $2.5$th and $97.5$th percentiles, together with
the fraction of resamples in which the reversal is present, $P({\rm flip})$.
The ensemble sizes used are listed in Table~\ref{tab:params}.

The numerical parameters of all three settings (lattice sizes, time steps, coarse-graining intervals, window widths and dictionary sizes) are collected in
Table~\ref{tab:params} of Appendix~\ref{app:numerics}, together with the
matrix-free implementation used at $N = 20$. Two analysis conventions that
affect the conclusions quantitatively are stated separately in
Appendix~\ref{app:conventions}.

\section{Results I: property currents in a complete interferometer}
\label{sec:resI}

The Cheshire-cat assignment itself is established: it was proposed in
Ref.~\cite{Aharonov2013cheshire}, realized with neutrons in
Ref.~\cite{Denkmayr2014}, and its dynamical version, a current of a property
unaccompanied by a current of the particle, was put forward in
Refs.~\cite{Aharonov2015current, Aharonov2021dynamical}. Reproducing it is
therefore not the point of this section. What we add is that the structure is
not imposed: the conditioning is specified only at the two ports, nothing is
engineered inside the arms, and the separation appears there on its own. We
then state it dynamically, as a pair of conservation laws for two
constructively defined currents rather than as an assignment at a single
instant, and make it operational with a response experiment that locates the
property, quantifies what detecting it costs, and turns the whole effect into a
property of an extracted generator. The last of these is what carries over to
the many-body sections: the same machinery applied to a chaotic chain is what
Secs.~\ref{sec:resII} and \ref{sec:resIII} use.

\subsection{Port-defined conditioning}
\label{sec:resI-setup}

We work on a two-rail ladder of $m=40$ sites per rail with an internal
spin-$1/2$, so $\dim \mathcal{H} = 160$. Writing $\ket{r,k}$ for site
$k = 1,\dots,m$ of rail $r \in \{U, D\}$, $\Pi_{r} = \sum_{k}
\ket{r,k}\bra{r,k}$ for the rail projectors and $n_{i}$ for the site
projectors, the Hamiltonian is
\begin{align}
  H(t) &= -J \sum_{r}\sum_{k=1}^{m-1}
          \left(\ket{r,k}\bra{r,k\!+\!1} + \text{h.c.}\right) \otimes \openone
          \nonumber\\
       &\quad + \lambda(t) \sum_{k=1}^{m}
          \left(-i\ket{U,k}\bra{D,k} + i\ket{D,k}\bra{U,k}\right)
          \otimes \openone \nonumber\\
       &\quad + \chi(t)\, \Pi_{D} \otimes \sigma_{y}
          \; + \; B(t)\, \Pi_{F} \otimes \sigma_{z} ,
  \label{eq:H-interf}
\end{align}
with open boundaries along each rail. The first line is longitudinal hopping;
the second is a spatially uniform rung coupling carrying a $\pi/2$ Peierls
phase, i.e.\ $\sigma_{y}$ in the rail (path) degree of freedom, which acts as a
beam splitter when pulsed; the third rotates the spin on rail D; and the fourth
is a Zeeman field along $z$, of amplitude $B$ and acting only on the
sites selected by the projector $\Pi_{F}$, which we use in
Sec.~\ref{sec:resI-response} as a weak local probe of the property assignment
and which is absent ($B = 0$) in the present subsection. Since
$\hbar = 1$ and the magnetic moment is absorbed into $B$, the latter is
a rate: it splits the two spin states on the affected sites by $2B$. The pulse
envelopes $\lambda(t)$ and $\chi(t)$ are rectangular and their areas and timings
are listed in Appendix~\ref{app:numerics}; during the interval between the
two beam-splitter pulses the rails are referred to as arms. The pre-selection injects a wave
packet into the input port of rail U with spin up; the post-selection detects
spin up at the output port of rail D. The two ports are the two ends of the
ladder: the packet is launched at site $x_{0}=6$ of rail U at $t=0$ and, with
group velocity $2J$, arrives near site $36$ of rail D at $t=T=16$, which is
where the detector sits. Both are product states of a single rail:
states that can be prepared and measured, not engineered inside the arms. The
post-selected state is the freely propagated image of the same packet shape on
rail D, i.e.\ a mode-matched detector rather than a single site; matching the
detector to the emitted mode
is what makes the overlap exactly $1/2$ rather than merely of that order, and
none of the conclusions below depend on it.

\begin{figure*}[!tb]
  \centering
  \includegraphics[width=\textwidth]{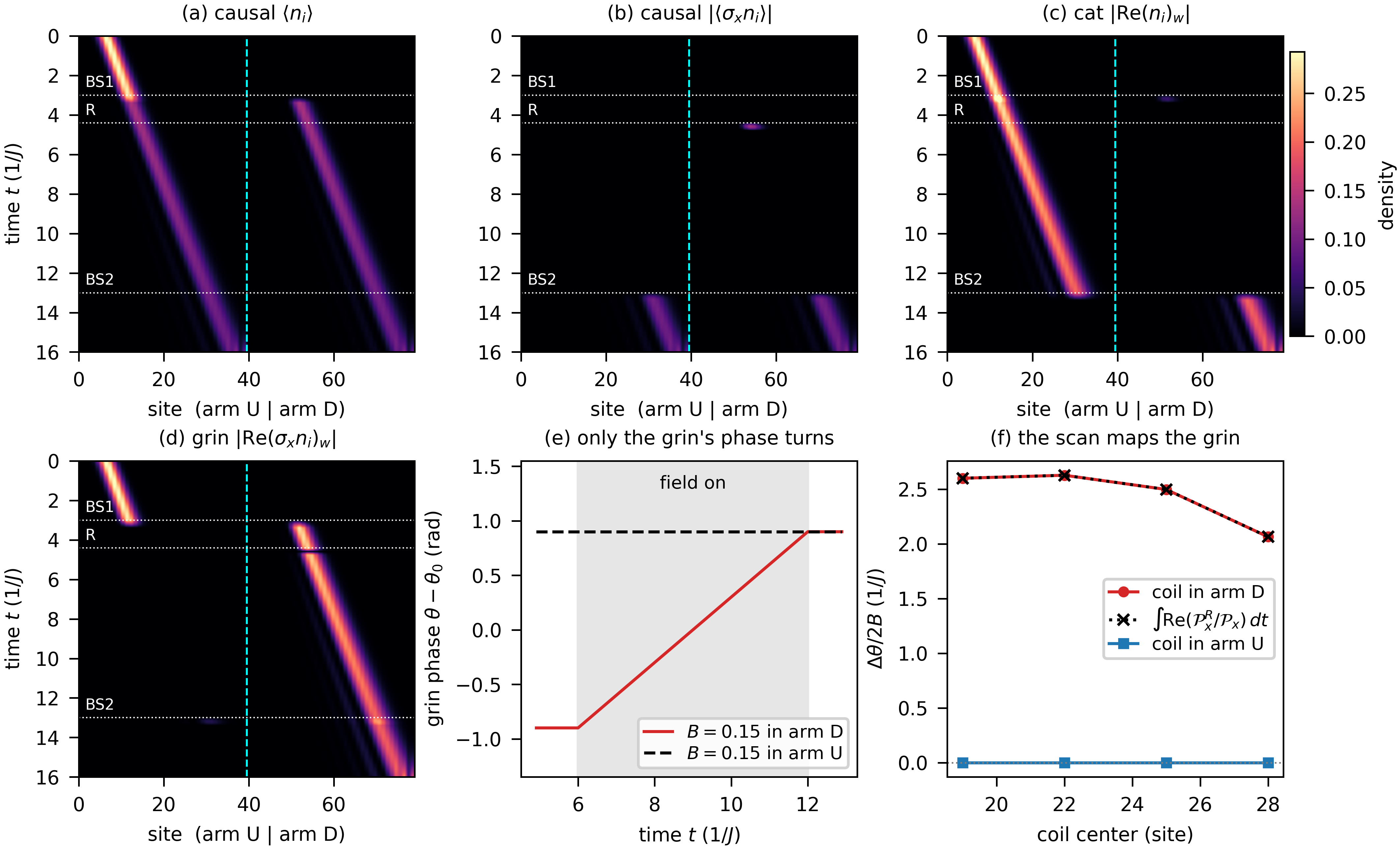}
  \caption{\textbf{Property currents and the response experiment in a
    port-defined interferometer.}
    (a) Causal particle density: the packet splits 50/50 at the first beam
    splitter. (b) Causal polarization density, which vanishes inside the arms.
    (c) Conditioned particle (``cat'') density, confined to arm U.
    (d) Conditioned polarization (``grin'') density, confined to arm D.
    (e) Response: with a field $B\,\sigma_{z}$ in arm D the grin's
    phase advances at exactly $2B$ (in radians per unit time) while the
    field is on (gray band),
    and is offset by $-B W$ \emph{before} the field is switched on, $W$ being
    the width of the field window; the same field in arm U produces no rotation
    at all (dashed), only a constant displacement $+BW$. Both constants are the
    field acting through the backward-propagated post-selection, which is why
    they are already present before the field is on; the rotation, i.e.\ the
    slope, is what distinguishes the two arms.
    In (a)--(d) the cyan dashed line marks the rail U/rail D boundary, the
    white dotted lines the beam-splitter (BS1, BS2) and rotator (R) pulses, and
    the shared color scale is the site-resolved density normalized to
    $\sum_{i} = 1$; because the conditioned densities are signed, (c) and (d)
    show the modulus of the real part, so the node interrupting the grin at the
    rotator in (d) is a sign reversal of $\mathrm{Re}\wv{\sigma_{x}n_{i}}$
    rather than a gap. All quoted values refer to the arm interval between R and
    BS2.
    (f) Coil scan: the accumulated phase reproduces
    $2B \int \mathrm{Re}(\mathcal{P}_{x}^{R}/\mathcal{P}_{x})\,dt$, with $\mathcal{P}_{x}$ the total
    grin amplitude and $\mathcal{P}_{x}^{R}$ its restriction to the coil region, and
    therefore maps the grin's spatial density; a coil in arm U rotates
    nothing. The ordinate of (f) is that integral, i.e.\ the dwell time of the
    grin within the coil, and therefore has the dimension of time.}
  \label{fig:interferometer}
\end{figure*}

Under this port-only specification the arm region develops the canonical
Cheshire-cat pair automatically, with
$\braket{\varphi}{\psi} = 1/2$ exactly.
Figure~\ref{fig:interferometer}(a)--(d) shows the result and contrasts it with
the causal description of the same experiment: conditioned on both ends, the
particle occupies arm U alone and the polarization arm D alone, whereas
conditioned on the past alone the particle is spread over both arms and no
polarization appears in either. The numbers behind those maps are as follows.
In the arms
$\wv{\Pi_{U}} = 1$ to $1.3\times10^{-14}$, with the particle's weak density on
rail D at most $1.1\times10^{-16}$: the particle is assigned entirely to
arm U. The $\sigma_{x}$
polarization is assigned entirely to arm D, its density in arm U vanishing
identically. The causal description gives the particle
half in each arm and no polarization anywhere in the arms
($1.1\times10^{-16}$).

Two structural facts underlie this result. First, the exactness of the
separation is
protected by a symmetry: between the active pulses $\sigma_{z}$ is conserved
in the arm region, so
$\ket{\psi}$ is purely spin-down on rail D while $\ket{\varphi}$ is purely
spin-up there, and the vanishing of $\wv{\Pi_{D}}$ and of the polarization in
arm U follows from spin orthogonality rather than from fine tuning. Second,
the weak polarization vector is null: site by site,
$\wv{\sigma_{y}} = -i \wv{\sigma_{x}}$ to $1.2\times 10^{-32}$, so
$\wv{\sigma_{x}}^{2} + \wv{\sigma_{y}}^{2} = 0$ and the ``property'' is a
single complex scalar rather than a two-component real vector.

\subsection{The response experiment}
\label{sec:resI-response}

To test whether the assignment is operational, we switch on the last term of
Eq.~\eqref{eq:H-interf} in the arm to which the polarization is assigned, i.e.\
$\Pi_{F} = \Pi_{D}$, so that every site of rail D feels the same field
$B\,\sigma_{z}$, over a time window of length $W = 6$ ($t \in [6,12]$,
entirely inside the arm region), and ask what
responds. The answers are exact rather than perturbative:

\begin{enumerate}
  \item The particle's weak density is unchanged to $2.2\times 10^{-16}$.
  \item The polarization's phase $\theta \equiv \arg \mathcal{P}_{x}$, with
        $\mathcal{P}_{x} = \sum_{i \in D}\wv{\sigma_{x} n_{i}}$, not to be
        confused with the overlap $g$ of Eq.~\eqref{eq:wv-def}, the total polarization
        amplitude in arm D, advances at rate exactly $2B$
        (measured $+0.30000$ for $B = 0.15$). The factor of two is a
        two-state effect: $\ket{\psi}$ and $\ket{\varphi}$ carry opposite
        $\sigma_{z}$ eigenvalues on rail D and accumulate the phase from both
        the past and the future boundary.
  \item Before the field is switched on, the phase is already offset by
        $-B W$; with the field in arm U the offset is $+B W$ and
        the rate is exactly zero ($-1.6\times10^{-17}$). The rate encodes the
        present field, the offset the future one.
  \item The response is invisible in the causal description
        ($3.1\times10^{-15}$) and in the post-selection probability $|g|^{2}$.
\end{enumerate}

In the language of Sec.~\ref{sec:method-currents}, the field enters as a source
in the \emph{property's own} conservation law,
\begin{equation}
  \frac{d}{dt}\wv{\sigma_{x} n_{i}}
    = -\nabla\!\cdot\! \wv{j^{\sigma_{x}}}
      - 2B \wv{\sigma_{y} n_{i}} ,
  \label{eq:grin-source}
\end{equation}
where $\wv{j^{\sigma_{x}}}$ is the polarization current, i.e.\
Eq.~\eqref{eq:current-def} with $S = \sigma_{x}$, and $\nabla\!\cdot\!$ the
lattice divergence $-\sum_{j}$ of Eq.~\eqref{eq:continuity};
verified to $1.9\times10^{-16}$ and localized where the property is. Restricting
$\Pi_{F}$ to a five-site coil ($B = 0.05$, centers scanned along the arm,
on for the whole arm interval) confirms the local statement
$d\theta/dt = 2B\,\mathrm{Re}(\mathcal{P}_{x}^{R}/\mathcal{P}_{x})$, where
$\mathcal{P}_{x}^{R}$ is the same sum restricted to the coil region, so that the accumulated
phase measures the property's spatial density
[Fig.~\ref{fig:interferometer}(f)]. The complementary statement also holds: a
coil in
arm D leaves the particle untouched to $6\times10^{-17}$, whereas a coil in
arm U scatters the particle (at the $10^{-2}$ level) yet does not rotate the
polarization.

\subsection{Generator-level statement and the scheme dichotomy}
\label{sec:resI-generator}

\begin{figure}[!tb]
  \centering
  \includegraphics[width=\columnwidth]{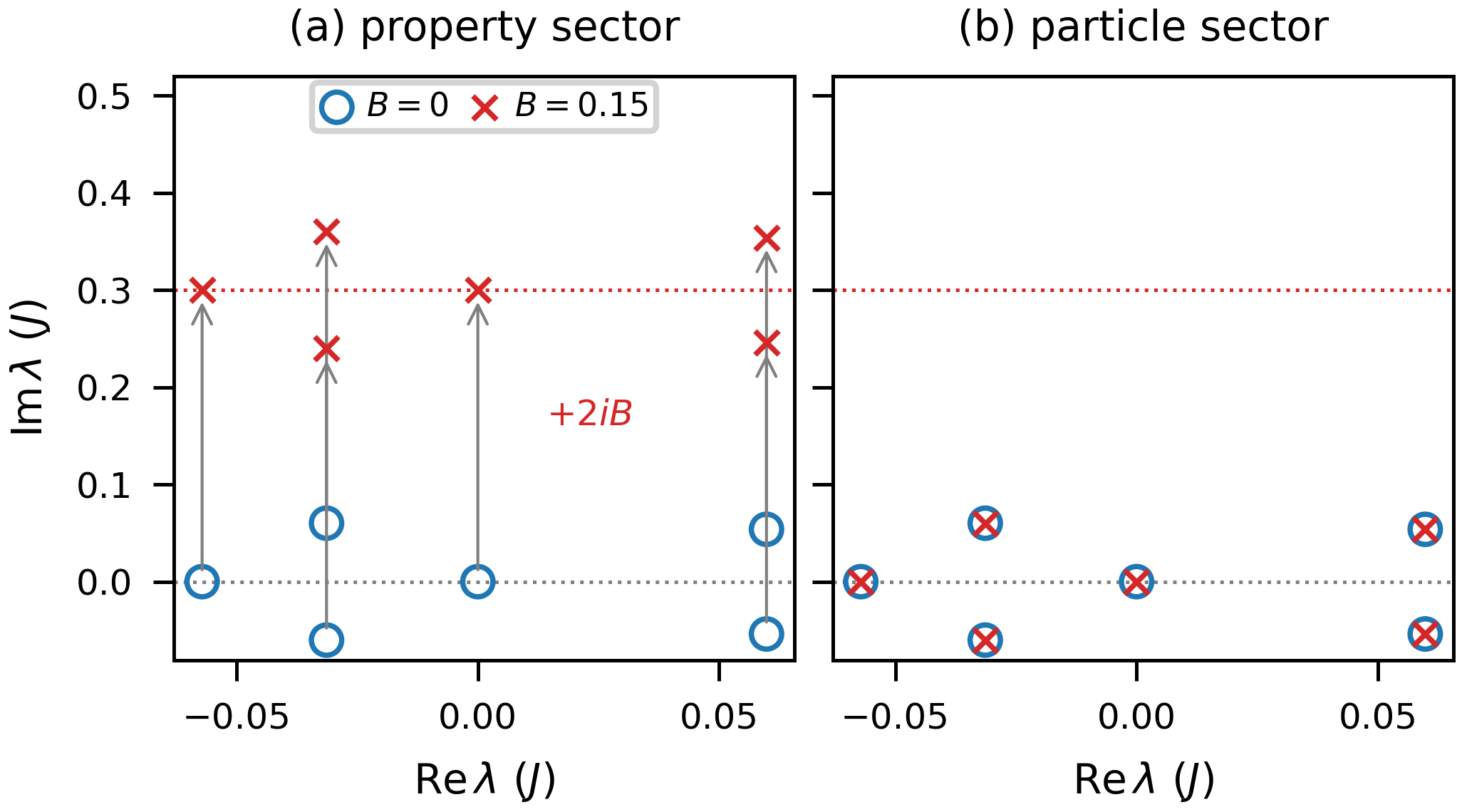}
  \caption{\textbf{The field enters the generator as a rigid spectral shift.}
    Eigenvalues of the reduced generator $L$ extracted from the spatial moments
    of each sector, without the field (circles) and with $B=0.15$ in arm D
    (crosses); arrows join eigenvalues matched to the prediction
    $\lambda \to \lambda + 2iB$. (a) In the
    property sector every eigenvalue moves vertically by the same amount, so
    that $\mathrm{Im}\,\lambda$ increases by $2B$ while $\mathrm{Re}\,\lambda$
    is untouched; this is the content of
    $L(B) = L_{0} + 2iB\openone$, and the dotted red line marks $2B = 0.30$.
    (b) In the particle sector the two sets coincide, the crosses sitting
    inside the circles. The deviations from exact rigidity,
    $2.4\times10^{-10}$ and $3.7\times10^{-11}$, are far below the marker size
    and are the conditioning floor of the regression rather than a physical
    effect.}
  \label{fig:spectra}
\end{figure}

Applying two-state gEDMD to spatial moments of the two sectors turns the
response into a statement about dynamics. The dictionary is the same for both
sectors and contains six elements, the spatial moments
$\sum_{i} x_{i}^{p}\, d_{i}(t)$ with $p = 0,\dots,5$, where $d_{i}$ is the
conditioned density on site $i$, namely $\wv{\sigma_{x}n_{i}}$ on rail D for
the property sector and $\wv{n_{i}}$ on rail U for the particle sector, and
$x_{i}$ is the site coordinate rescaled so that the rail spans $[-1,1]$. The derivatives are the exact ones of
Eq.~\eqref{eq:exact-deriv}, so no differencing interval enters here and the
comparison is between two generators rather than between two estimators; the
regression is the truncated least squares of
Sec.~\ref{sec:method-gedmd}, at the $10^{-9}$ threshold quoted there. The property sector's spectrum shifts
rigidly by $+2iB$ while
the particle sector's spectrum is unchanged [Fig.~\ref{fig:spectra}]:
$L(B) = L_{0} + 2iB\,\openone$. Every eigenvalue of the property sector moves
by the same purely imaginary amount, its real part untouched, and every
eigenvalue of the particle sector stays where it was; the residual deviations
from exact rigidity are $2.4\times10^{-10}$ and $3.7\times10^{-11}$, far below
the size of the plotted markers. The shift is exact rather than approximate,
and the argument is short. While the field is on, the entire property
trajectory equals
the zero-field trajectory times a common phase
$e^{2iB (t-t_{\rm mid})}$, $t_{\rm mid}$ being the midpoint of the field
window. A common phase leaves $\bm{X}\bm{X}^{\dagger}$, and hence the
singular-vector basis, invariant, so the regression returns the same generator up to the
derivative of that phase, which is $2iB$ times the identity. The argument holds
only where the phase relation does, so the fit is restricted to
$t \in [6.2, 11.8]$, inside the field window $[6, 12]$ by a margin of $0.2$ at
each end. The residual deviations ($10^{-11}$--$10^{-10}$) are the
conditioning floor of the least-squares regression, not a physical effect.

Reducing the dictionary to its $p=0$ element alone, which since $x_{i}^{0}=1$
is the total polarization amplitude
$\mathcal{P}_{x} = \sum_{i \in D}\wv{\sigma_{x}n_{i}}$ of
Sec.~\ref{sec:resI-response}, exposes the dichotomy that
motivates the rest of the paper (Table~\ref{tab:scheme}).

\begin{table}[b]
  \caption{\textbf{The differencing scheme manufactures friction.}
    Generator extracted from the single complex observable $\mathcal{P}_{x}$, for which
    $L$ is the scalar $\dot{\mathcal{P}}_{x} \approx L \mathcal{P}_{x}$; its real part is a decay
    rate (friction) and its imaginary part a precession rate, both in units of
    $J$. The two rows use the same trajectory and differ only in how
    $\dot{\mathcal{P}}_{x}$ is obtained: exactly from Eq.~\eqref{eq:exact-deriv}, or by a
    forward difference at $\dtcg = C\,dt = 0.4$ (Table~\ref{tab:params}). The
    friction in the second row is not a property of the trajectory; it scales
    with $\dtcg$ and vanishes with it (Appendix~\ref{app:conventions}).}
  \label{tab:scheme}
  \begin{ruledtabular}
  \begin{tabular}{lcc}
    estimator & $\mathrm{Re}\,L$ (friction) & $\mathrm{Im}\,L$ (precession) \\
    \hline
    exact derivative   & $-1.1\times10^{-17}$ & $+0.30000$ \\
    forward difference & $-1.77\times10^{-2}$ & $+0.29649$ \\
  \end{tabular}
  \end{ruledtabular}
\end{table}

The same data, differing only in how the derivative is taken, yield a physical
precession that is estimator-independent and an apparent friction that the
forward difference introduces. Its size is predicted exactly. For a pure
rotation $\mathcal{P}_{x} \propto e^{2iBt}$ the forward difference returns
$[e^{2iB\dtcg}-1]/\dtcg$, whose real part is
$-(2B)^{2}\dtcg/2 = -1.80\times10^{-2}$ at $\dtcg = 0.4$ and $2B = 0.30$,
matching the measured $-1.77\times10^{-2}$ to $1.5\%$. The apparent friction is
therefore proportional to the differencing interval and vanishes with it; the
value in Table~\ref{tab:scheme} is not to be compared directly with the
many-body numbers quoted later, which use $\dtcg = 0.04$.

\subsection{Robustness of the separation}
\label{sec:resI-robustness}

The statements above hold to machine precision in an idealized model, so it
remains to establish how much of that survives departures from the
idealization. Two scans address this: the separation is
fragile against breaking its
protecting symmetry, and robust against imperfections that respect it.

\begin{figure}[!tb]
  \centering
  \includegraphics[width=\columnwidth]{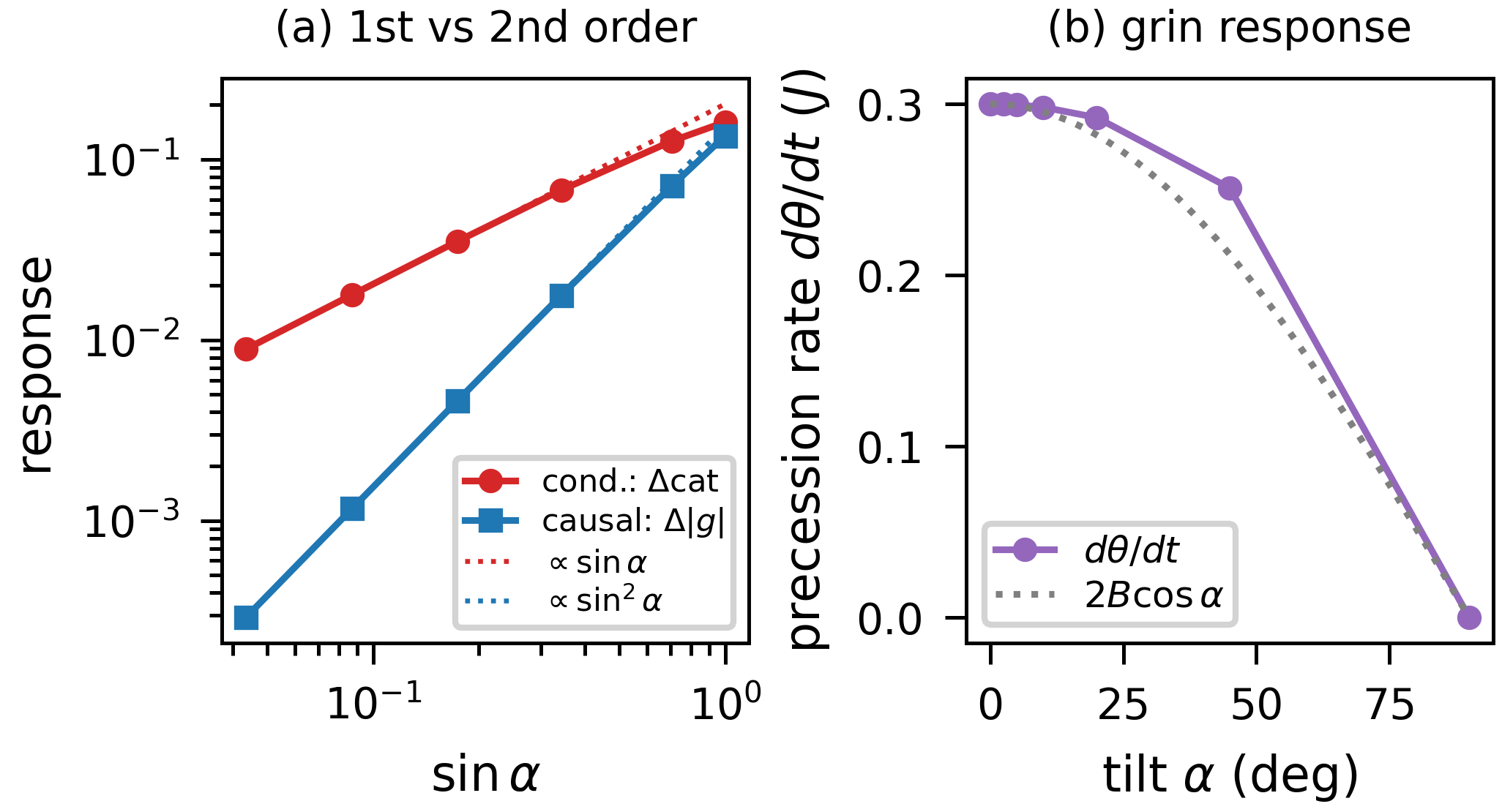}
  \caption{\textbf{Symmetry protection is first order in the conditioned
    description and second order in the causal record.}
    (a) Tilting the field out of the $\sigma_{z}$ axis by $\alpha$ moves the
    conditioned particle density as $\sin\alpha$, while the only causal
    quantity that responds, the post-selection probability through $|g|$, responds as $\sin^{2}\alpha$. (b) The grin's precession rate
    follows $2B\cos\alpha$.}
  \label{fig:robustI}
\end{figure}

Tilting the field out of the $\sigma_{z}$ axis by an angle $\alpha$ breaks the
protection, and it does so at first order
[Fig.~\ref{fig:robustI}(a)]. Writing
$\Delta(\text{cat}) \equiv \max_{i}\left|\wv{n_{i}}(\alpha) -
\wv{n_{i}}(0)\right|$ for the largest change of the conditioned particle
density over the arm region, and $\Delta|g|$ for the corresponding change in
the overlap, fitting the scaling gives
$\Delta(\text{cat}) \propto B^{0.99}\sin^{1.00}\!\alpha$, so the
particle's assignment is not robust in any protected sense once the perturbation
fails to commute with $\sigma_{z}$; the exactness reported in
Sec.~\ref{sec:resI-response} is a consequence of the symmetry alone.
The comparison with the causal description makes this quantitative. The causal particle density is \emph{exactly} invariant for every
$\alpha$, because the field is diagonal in position and commutes with $n_{i}$;
the only causal quantity that responds at all is the post-selection probability,
and it responds at second order, $\Delta|g| \propto \sin^{1.97}\!\alpha$. The
conditioned description is therefore one order in the perturbation more
sensitive than the statistics an experimenter records. The grin, meanwhile, never leaks into arm~U for
any field direction, the spin purity of rail~U being untouched, and the
precession rate follows $2B\cos\alpha$ to within $0.003$ at
$\alpha = 10^{\circ}$ and $0.011$ at $20^{\circ}$
[Fig.~\ref{fig:robustI}(b)].

\begin{figure}[!tb]
  \centering
  \includegraphics[width=\columnwidth]{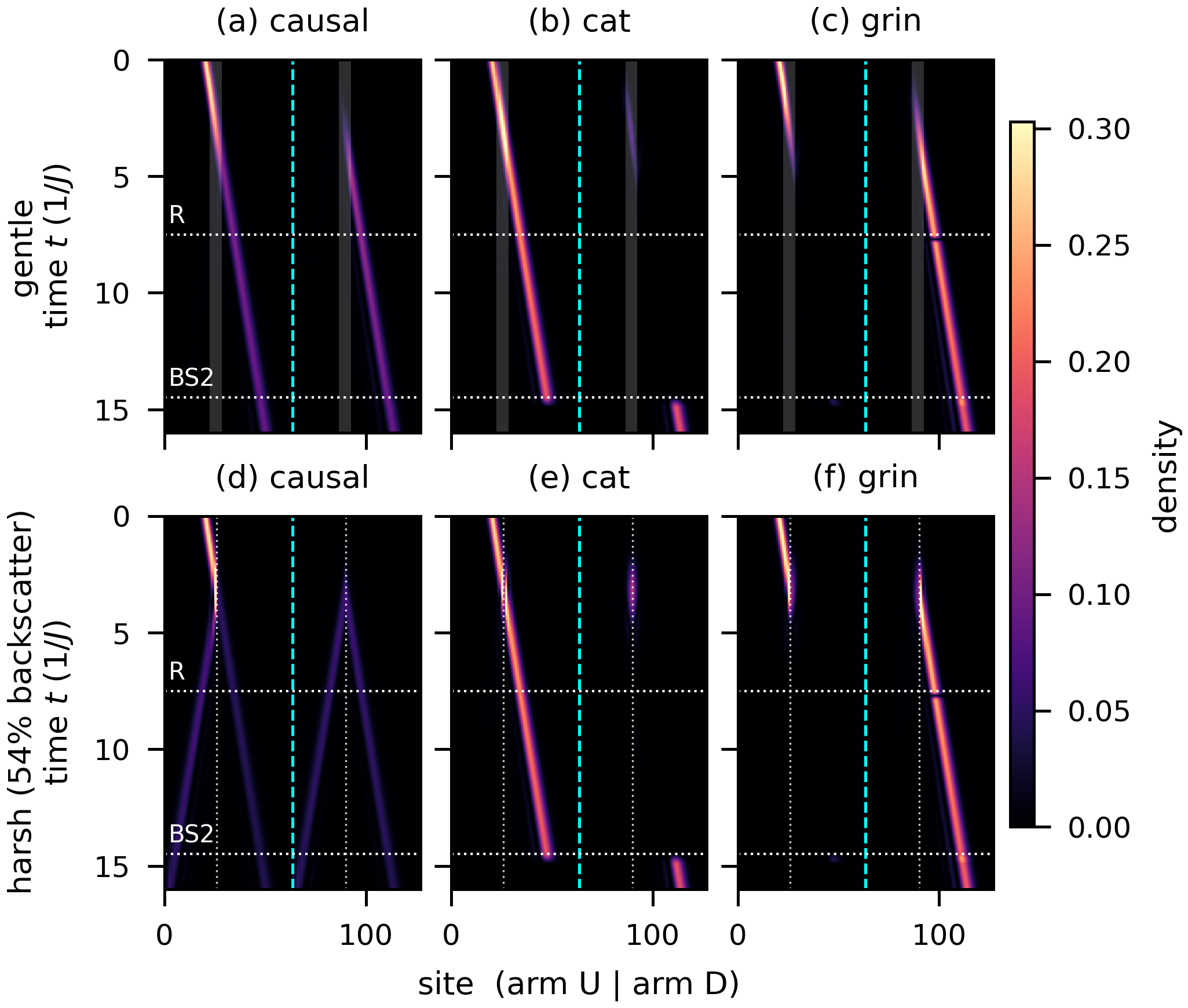}
  \caption{\textbf{The separation survives a lossy junction.}
    Top row (a)--(c): a gentle localized coupler ($2\%$ backscatter). Bottom
    row (d)--(f): a
    single-site coupler with coupling equal to the group velocity, which
    backscatters $54\%$ of the incident probability and produces the large reflected
    branch visible in the causal density (d). The conditioned cat (e) and grin
    (f) nevertheless recover the ideal single-branch structure, with residual
    leakage at the $10^{-3}$ level, three orders below the signal and hence
    not resolved on the linear color scale. As in
    Fig.~\ref{fig:interferometer}, the cyan dashed line is the rail boundary and
    the conditioned panels show the modulus of the real part; the single color
    bar applies to all six panels, so the two rows may be compared directly, and
    it saturates at the $99.9$th percentile of the six maps (the incident peak
    of (d) exceeds it). The horizontal white dotted lines mark the spin rotator
    (R, here at $t=7.5$) and the output beam splitter (BS2); the node
    interrupting the grin
    in (c) and (f) is the rotator reversing the sign of
    $\mathrm{Re}\wv{\sigma_{x}n_{i}}$, which the modulus renders as a zero.
    Quantities are quoted for the arm interval between R and BS2. Unlike the
    input splitter of Fig.~\ref{fig:interferometer}, the junction studied here
    is \emph{static and spatially localized}, so it has no pulse time to mark;
    its position is indicated instead by the vertical white markers (a six-site
    band in the top row, the single site $26$ in the bottom row, in both rails),
    and the splitting occurs where the packet's worldline crosses them.
    This scan uses $m = 64$ sites per rail so that the backscattered branch does
    not return from the input edge within the run.}
  \label{fig:yjunction}
\end{figure}

Against imperfections that respect the symmetry, the separation is
robust. We replaced the pulsed, spatially uniform beam splitter of
Eq.~\eqref{eq:H-interf} by a static localized junction: the rung term is
retained on a few sites only and left on at all times, so that the splitting
happens where the packet crosses it rather than when a pulse fires. Spreading
it over six rungs, with the coupling tuned for a $50$/$50$ split, backscatters
only $2\%$ of the incident probability (the ``gentle'' coupler); concentrating it on a
single rung with coupling equal to the group velocity backscatters $54\%$ (the
``harsh'' one) and reduces the
post-selection overlap to $0.33$ (against $1/2$ for the uniform splitter and
$0.48$ for the gentle coupler). Causally this is a severe
distortion: the density map shows a large reflected branch traveling backwards
[Fig.~\ref{fig:yjunction}]. The conditioned description, however, recovers
the ideal single-branch structure, with the residual leakage of the cat into
arm~D and of the grin into arm~U at the $10^{-3}$ level, three orders below the
signal, and the response experiment unaffected ($d\theta/dt = 0.29997$ against
the ideal $0.300$). Moreover, the leakage does not scale with the loss: going
from the gentle coupler ($2\%$ backscatter) to the harsh one ($54\%$) multiplies
the backscattered fraction by $27$ but the leakage only by $2$. The residual is
not caused by the loss at all but by the temporal overlap of the elements, the dispersive tail of the packet still sitting in the always-on coupler when the spin rotator fires, and it falls monotonically as the rotator is delayed. What
protects the separation is $\sigma_{z}$ conservation, not spatial perfection of
the interferometer.

\section{Results II: the reflection involution and a two-layer arrow of time}
\label{sec:resII}

\subsection{Setting}
\label{sec:resII-setting}

We now take the friction itself as the object of study, in a chaotic XXZ chain
of $N=8$ spins with open boundaries in space,
\begin{equation}
  \begin{split}
    H &= \sum_{i=1}^{N-1}\left[ J\left(X_{i}X_{i+1} + Y_{i}Y_{i+1}\right)
          + \Delta\, Z_{i}Z_{i+1} \right] \\
      &\quad + J_{2}\sum_{i=1}^{N-2} Z_{i}Z_{i+2} ,
  \end{split}
  \label{eq:H-xxz}
\end{equation}
where $X_{i}, Y_{i}, Z_{i}$ are the Pauli operators on site $i$, the exchange
is $J = \Delta = 1$ and $J_{2} = 0.5$, and the Hilbert-space dimension is
$D = 2^{N}$. The next-nearest-neighbor coupling is
of $ZZ$ type only, which breaks the integrability of the bare XXZ chain while
keeping $H$ real symmetric in the computational basis, the hypothesis on which
the involution of Sec.~\ref{sec:resII-involution} rests. Such chains thermalize in the
sense of the eigenstate thermalization
hypothesis~\cite{Deutsch1991, Srednicki1994, Rigol2008} and support
hydrodynamic transport of the conserved
magnetization~\cite{Bertini2021}; it is the friction of that hydrodynamics
which our companion study extracted for a causal
observer~\cite{Saito2026sister}.

The collective coordinates we follow are a long-wavelength magnetization
amplitude and its velocity,
\begin{equation}
  a = \sum_{i} p_{i} Z_{i}, \qquad b = i[H, a] = \dot a ,
  \label{eq:mode-ops}
\end{equation}
with profile $p_{i} = \cos[\pi k i/(N-1)]$, $k=1$ unless stated otherwise.
Their weighted mean trajectories are fitted, on each sliding window, to a
damped harmonic law
\begin{equation}
  \dot{\bar b}_{w} = -\Omega^{2}\, \bar a_{w} - \gamma_{m}\, \bar b_{w} ,
  \label{eq:mode-fit}
\end{equation}
This is the regression of Sec.~\ref{sec:method-gedmd} with the two-element
dictionary $\{a, b\}$: Eq.~\eqref{eq:mode-fit} is the row of $L$ that returns
$\dot b$, and with only two dictionary elements the singular-value truncation
never activates, so the fit reduces to an ordinary least squares in the two
coefficients (taken over the real and imaginary parts jointly, since the weak
values are complex). Those coefficients are the two quantities we
report at this layer: $\Omega^{2}$, the restoring rate, which has units of
$J^{2}$ and plays the role of a squared oscillation frequency for the mode, and
$\gamma_{m}$, the mode-layer friction, in units of $J$. We write $\Omega^{2}$
rather than $\Omega$ throughout because the fit determines that combination
directly and it need not be positive. Both are functions of the window center
$t$, which is what makes statements about their behavior across the midpoint
meaningful.

Both boundaries are drawn as in Eq.~\eqref{eq:seeds}, with $w$ independent of
$v$; this is the class on which the theorem below is proved.
The modulation operator is $\Lambda = a$: the boundaries excite precisely the
mode whose effective dynamics we then fit, and the scalar
$\varepsilon$ controls how strongly they do so. We use $\varepsilon = 0.35$
throughout except where it is scanned in
Sec.~\ref{sec:resII-robustness}; larger $\varepsilon$ means a more strongly
conditioned and, since $|g|$ grows with it, brighter ensemble. Unlike
Sec.~\ref{sec:resI}, every quantity reported below is an ensemble average in
the sense of Eq.~\eqref{eq:selfaverage}: the figures of this section use
$M = 160$ reflection-paired samples, and the robustness scans of
Sec.~\ref{sec:resII-robustness} use $M = 200$ per point, or up to $400$ where
$M$ itself is varied.

A natural expectation is that conditioning on
the future reverses the arrow of time. We show that it does so only at one
level of description, and that the constraint follows from an exact symmetry
of the ensemble.

\subsection{The reflection involution}
\label{sec:resII-involution}

Consider the map
\begin{equation}
  R : (\ket{\psi_{0}}, \ket{\varphi_{T}}) \longmapsto
      (\ket{\varphi_{T}^{*}}, \ket{\psi_{0}^{*}}) .
  \label{eq:involution}
\end{equation}
For real symmetric $H$ and a conjugation-invariant i.i.d.\ measure, $R$ is a
measure-preserving involution of the $|g|^{2}$-weighted ensemble
(Appendix~\ref{app:proof}), and for any
observable with a definite reflection signature $O^{*} = \epsilon_{O} O$ it
implies, sample by sample,
\begin{equation}
  g' = g, \qquad O'_{w}(t) = \epsilon_{O}\, O_{w}(T-t) ,
  \label{eq:sample-identity}
\end{equation}
where a prime denotes the quantity evaluated on the reflection partner
$R(\ket{\psi_{0}}, \ket{\varphi_{T}})$.
Here $\epsilon_{O} = +1$ for time-even observables (the Pauli-$Z$ densities
$Z_{i}$ and $Z_{i}Z_{j}$) and $-1$ for time-odd ones (currents). We verify
Eq.~\eqref{eq:sample-identity} to $10^{-11}$ over a $29$-dimensional
hydrodynamic dictionary [Fig.~\ref{fig:bridge}(a)].

Being a sum of $Z$ densities, the amplitude $a$ of Eq.~\eqref{eq:mode-ops} is
time-even, while its time
derivative $b$ is time-odd; Eq.~\eqref{eq:sample-identity} therefore predicts
that the $|g|^{2}$-weighted ensemble means of Eq.~\eqref{eq:selfaverage}, written $\bar a_{w}$ and $\bar b_{w}$, are respectively symmetric and antisymmetric about $t=T/2$, which is
what Fig.~\ref{fig:bridge}(a) shows.

\subsection{Unique decomposition of the friction}
\label{sec:resII-decomposition}

\begin{figure}[!tb]
  \centering
  \includegraphics[width=\columnwidth]{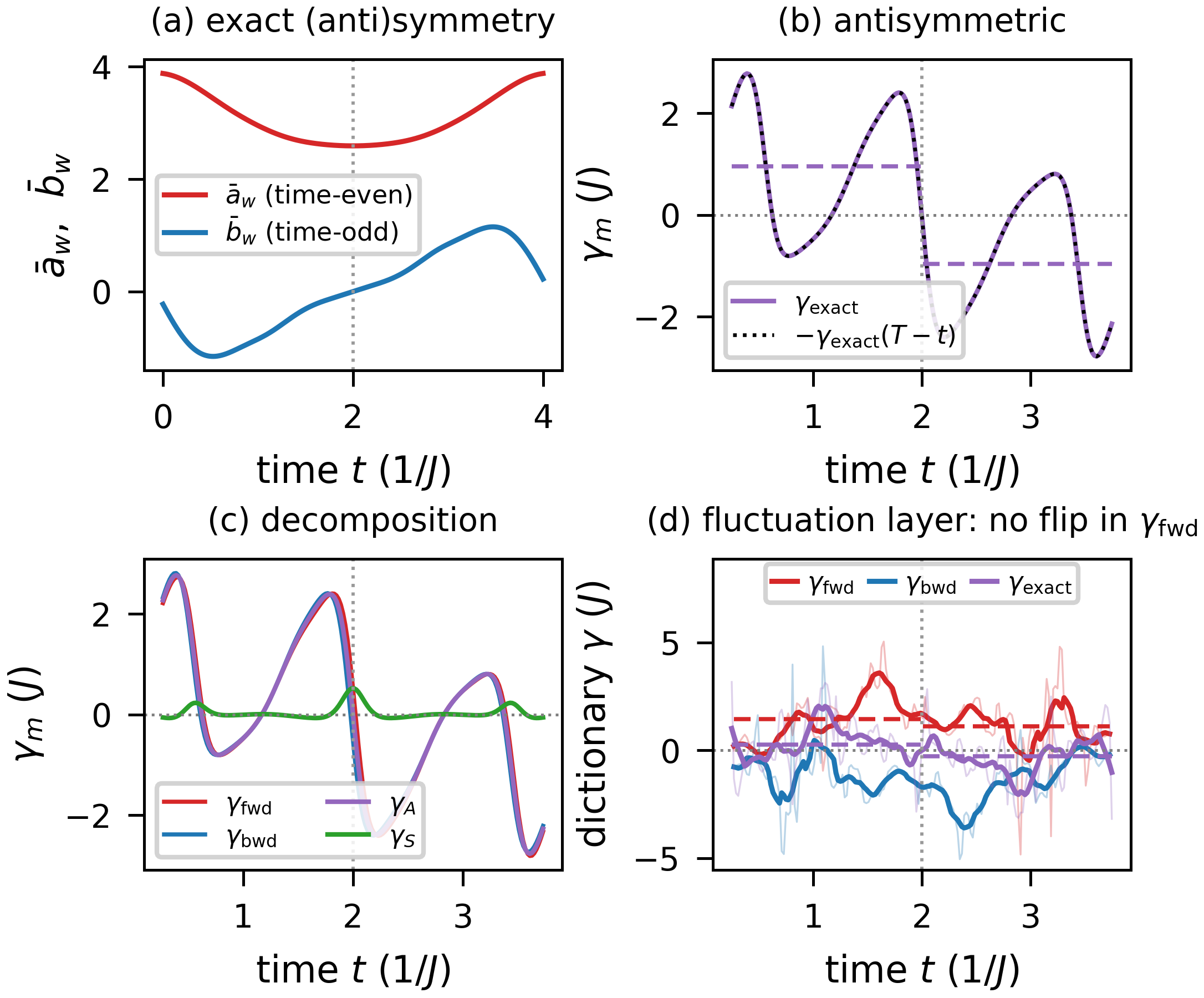}
  \caption{\textbf{The reflection involution and the two layers.}
    (a) On a reflection-paired ensemble the weighted means are exactly
    symmetric (time-even) or antisymmetric (time-odd); real parts are shown,
    and note that $\bar a_{w}$ is a dimensionless amplitude while
    $\bar b_{w} = \dot{\bar a}_{w}$ is a rate (units of $J$).
    (b) Mode-layer friction from the exact derivative. The dotted curve is
    $-\gamma_{\rm exact}(T-t)$, so their coincidence is the pointwise
    antisymmetry; the dashed levels are the half-window averages, whose sign
    reversal is the coarse-grained statement discussed in the text.
    (c) Decomposition at the mode layer, with the one-sided estimators taken at
    $\dtcg = 0.04$ here and in (d). That $\gfwd$, $\gbwd$ and $\gA$ nearly
    coincide is the result rather than a redundancy: their difference is the
    scheme part $\gS$, negligible here.
    (d) The same three estimators applied to the full dictionary, i.e.\ the
    fluctuation layer, where the situation is reversed and $\gbwd$ is the mirror
    $-\gfwd(T-t)$. Faint traces are the raw window fits, heavy ones an
    $11$-point running mean, and the dashed levels the half-window averages, in
    red for $\gfwd$, which is positive on both sides of the midpoint, and in
    purple for $\gamma_{\rm exact}$, which is not: the exact-derivative friction
    at this layer is antisymmetric to $9.6\times10^{-13}$ and averages
    $\pm 0.274$, so what the differencing suppresses is a reversal present in the
    exact generator (Sec.~\ref{sec:resII-twolayer}). Plotting the estimators
    together is what makes $\gA$ and $\gS$ separately visible. Here and in
    Figs.~\ref{fig:robustII-eps}--\ref{fig:q20} the vertical dotted line marks
    the midpoint $t = T/2$ and the horizontal dotted line zero friction.}
  \label{fig:bridge}
\end{figure}

Window-fitting estimators are linear maps of the data and therefore equivariant
under $R$, which exchanges forward and backward differences
(Appendix~\ref{app:proof}). On a
reflection-invariant ensemble this gives the exact relations
\begin{equation}
  \gfwd(t) = -\gbwd(T-t), \qquad
  \gamma_{\rm exact}(t) = -\gamma_{\rm exact}(T-t),
  \label{eq:equivariance}
\end{equation}
with $\gamma_{\rm exact}$ the same window fit supplied with the exact
derivative~\eqref{eq:exact-deriv}, verified to $2\times10^{-13}$
[Fig.~\ref{fig:bridge}(b)], together with $\Omega^{2}(t) = \Omega^{2}(T-t)$:
the restoring rate of the mode fit~\eqref{eq:mode-fit} carries no time
asymmetry. Consequently every measured friction splits
uniquely as in Eq.~\eqref{eq:decomp-intro}, with
\begin{equation}
  \gA = \tfrac{1}{2}\left(\gfwd + \gbwd\right), \qquad
  \gS = \tfrac{1}{2}\left(\gfwd - \gbwd\right),
  \label{eq:decomp}
\end{equation}
where $\gA(t) = -\gA(T-t)$ agrees with the exact-derivative friction to
$1.2\%$ and is fixed by the boundary conditions, while $\gS(t) = +\gS(T-t)$ is
the coarse-graining artifact of Ref.~\cite{Saito2026sister} and cannot
reverse across the midpoint [Fig.~\ref{fig:bridge}(c)].

\subsection{The two layers}
\label{sec:resII-twolayer}

Which component dominates depends on the level of description; this is what
the term ``two-layer arrow of time'' refers to. The two layers are deliberately
different descriptions of the same ensemble rather than two estimates of one
quantity, and it is worth saying why they can nevertheless be set against each
other. The decomposition is defined separately within each layer, and the
equivariance that produces it, Lemma~3 of Appendix~\ref{app:proof}, requires
only that the window fit be a fixed linear-algebraic function of the trajectory
segment it is given. Both fits qualify, so
$\gfwd(t) = -\gbwd(T-t)$ holds at each layer on its own; we verify it to
$2\times10^{-13}$ at the mode layer and $1\times10^{-12}$ at the dictionary
layer. What we then compare across layers is not the friction itself, whose
value has no reason to agree between a two-variable fit and a $29$-dimensional
regression, but which of the two components of the same decomposition
dominates.

At the level of a coherent hydrodynamic mode, i.e.\ from the fit
\eqref{eq:mode-fit} to the weighted mean trajectory of the amplitude $a$ and
velocity $b$ of Eq.~\eqref{eq:mode-ops}, the antisymmetric component dominates, and the friction reverses sign at the midpoint of the conditioning
window: averaged over the first half of the window $\gamma_{m} = +0.96$ and over
the second $-0.96$ for
the exact estimator, with the forward difference giving $+0.98/-0.90$. These
are quoted on the full window grid, as in Fig.~\ref{fig:bridge} and
Table~\ref{tab:cgscan}; the ill-conditioned-window exclusion of
Appendix~\ref{app:conventions} lowers them by $7$--$8\%$ at this size
($\pm0.96 \to \pm0.89$) and changes nothing else. The
reversal is
therefore estimator-independent and is inherited from the boundary conditions,
that is, from the pre- and post-selection.
Quantitatively, the forward and backward estimators track each other pointwise
to within $37\%$ of the oscillation amplitude, so at this layer
$\gS$ averages to only $+0.04$ against $\gA = +0.94$: the friction one measures
is the physical component to within $4\%$
[Fig.~\ref{fig:bridge}(c)].
Two qualifications belong with these numbers. They are averages: because the
sliding window is shorter than the period of the mode, the fitted
$\gamma_{m}(t)$ oscillates about them with roughly three times their amplitude
and is transiently of the opposite sign for about a third of each half
[Fig.~\ref{fig:bridge}(b)]. What holds pointwise, and to $2\times10^{-13}$, is
the antisymmetry $\gamma_{m}(t) = -\gamma_{m}(T-t)$ of
Eq.~\eqref{eq:equivariance}; the sign reversal of the half-window averages is
its coarse-grained consequence. At $N=20$ the same oscillation is $16$ times
the average and we therefore report the convention-insensitive reflection
residuals alongside (Appendix~\ref{app:conventions}).

At the fluctuation layer the situation is inverted. This layer is the same
regression with the full $29$-dimensional dictionary in place of $\{a, b\}$,
and with one further difference: instead of averaging the samples first and
fitting the single mean trajectory, we stack the individual sample
trajectories side by side, each weighted by $|g|$, and regress them together,
so that the least-squares residual carries the $|g|^{2}$ weight of
Eq.~\eqref{eq:selfaverage} while the fluctuations about the mean are retained
rather than averaged away. That is what makes it a different layer of
description and not merely a larger dictionary; we use ``fluctuation layer''
and ``dictionary layer'' interchangeably for it. The
antisymmetric component is subdominant, $|\gA|/\gS = 0.135$, so
$\gfwd > 0$ in both halves and the reversal is \emph{forbidden}
[Fig.~\ref{fig:bridge}(d)]. The sharp
criterion is simply $\gS > |\gA|$, satisfied here with a factor of $7$ to
spare.

The comparison that fixes the scope of this statement is with the
exact-derivative estimator at the same layer. Supplied with
$\dot X = i\wv{[H,X]}$ instead of a finite difference, the dictionary
regression returns a friction that is antisymmetric about the midpoint to
$9.6\times10^{-13}$, with half-window averages $+0.274$ and $-0.274$
[the exact curve in Fig.~\ref{fig:bridge}(d)]. The fluctuation layer therefore
does not lack a reversing component; the reversal is buried by $\gS$, which
the differencing injects. What distinguishes the two layers is how much of that
scheme component each one receives: at the differencing resolution used here
$\gS = +0.038$ at the mode layer against $+1.278$ at the dictionary layer, a
factor of $34$, while the antisymmetric parts differ by only a factor of $5$
($0.94$ against $0.17$). The no-flip statement is thus a statement about a
description at a fixed inference resolution, not a resolution-independent
property of the fluctuation layer, and $\Delta t_{\rm cg}$ has to be quoted
with it. Appendix~\ref{app:conventions} gives the dependence: $\gS$ is
proportional to $\Delta t_{\rm cg}$ to within $4\%$ over a factor of eight in
resolution, so it vanishes with the coarse-graining interval while $\gA$ does
not, and the margin is set by the estimator as much as by the
system. This is the pre-/post-selected counterpart of the result of
Ref.~\cite{Saito2026sister}, that the friction extracted from a causal
description is a function of the observer's temporal resolution and carries no
net dissipation in the exact-derivative limit.

Because $\gA$ and $\gS$ are computed from the same data as
$(\gfwd \pm \gbwd)/2$, the boundary-condition arrow and the inference arrow are
separable by measurement. This separability is the central methodological
result of the paper.

\subsection{Robustness and the role of the conditioning class}
\label{sec:resII-robustness}

\begin{figure}[!tb]
  \centering
  \includegraphics[width=\columnwidth]{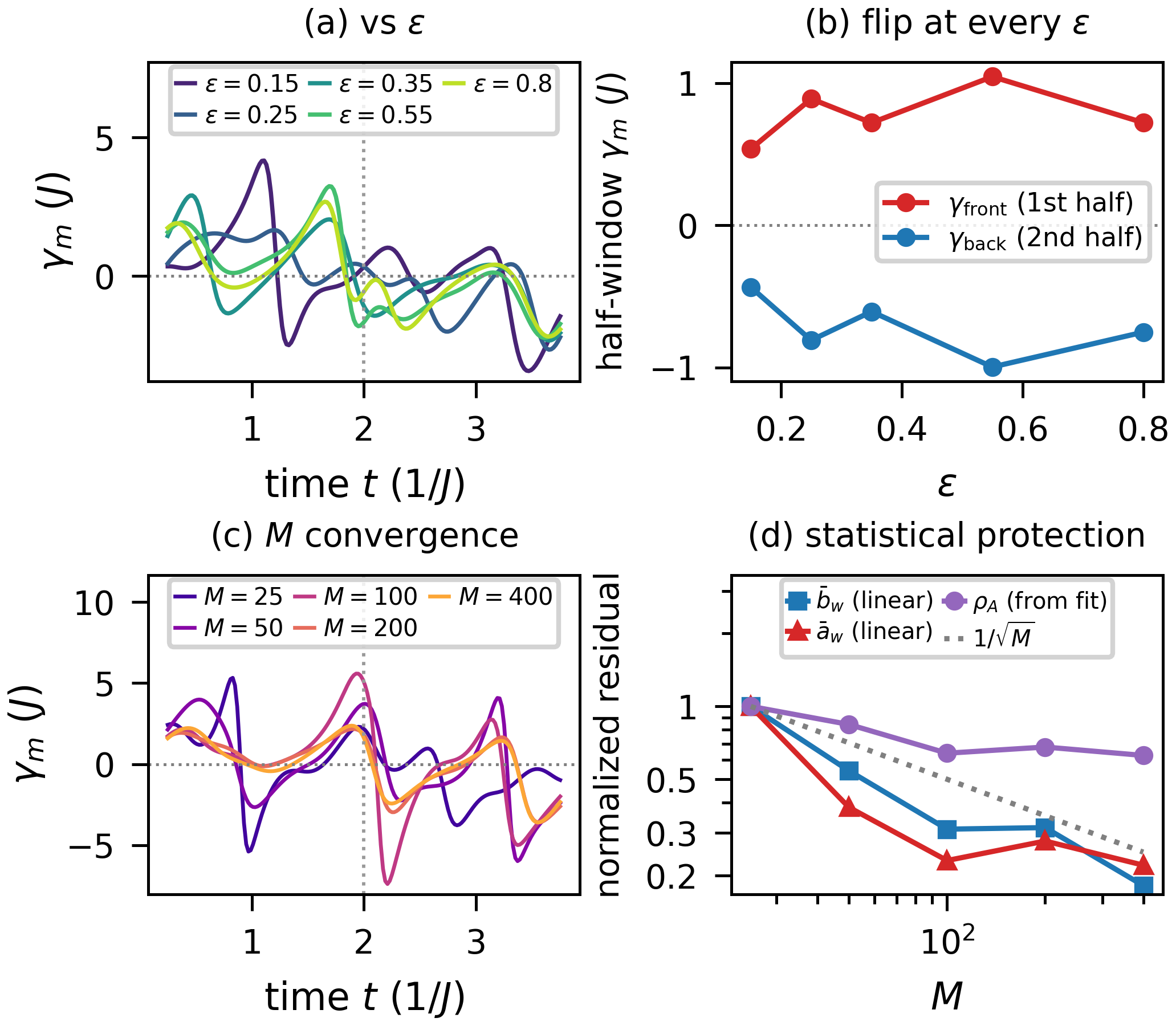}
  \caption{\textbf{Robustness in conditioning strength and ensemble size.}
    (a) Mode-layer friction $\gamma_{m}(t)$ for five conditioning strengths
    $\varepsilon$. As in Fig.~\ref{fig:bridge}(b) each curve oscillates about
    its half-window averages and is transiently of either sign, so the reversal
    is read off from those averages rather than pointwise: (b) shows them,
    $\gamma_{\rm front}$ and $\gamma_{\rm back}$ versus $\varepsilon$, keeping
    opposite signs at every $\varepsilon$. These two are the averages of
    $\gamma_{m}(t)$ over the windows centered in the first and second half of
    the conditioning window; despite the similar names they have nothing to do
    with the forward and backward estimators $\gfwd$ and $\gbwd$ of
    Fig.~\ref{fig:bridge}(c), which are not used anywhere in this figure.
    (c) The same curves at fixed $\varepsilon = 0.35$ for increasing ensemble
    size $M$,
    showing convergence. (d) How the predicted symmetries emerge with ensemble
    size. Each curve is divided by its own value at $M=25$ so that all three
    start at unity and their slopes can be read against the single
    $1/\sqrt{M}$ guide (dotted); the underlying residuals differ by more than an
    order of magnitude and would otherwise not be comparable by eye. The
    quantities the theorem speaks about directly are the weighted
    means, and their reflection residuals, $|\bar a_{w}(t) - \bar a_{w}(T-t)|$
    and $|\bar b_{w}(t) + \bar b_{w}(T-t)|$ normalized by their maxima, fall at
    least as fast as $1/\sqrt{M}$: the protection is a property of the
    ensemble average and not of individual samples. The residual $\rho_{A}$ of
    the \emph{fitted} friction, Eq.~\eqref{eq:residuals}, follows them only to
    $M \simeq 100$ and then saturates near $0.25$. That floor comes from the
    fit, not from the ensemble: $\gamma_{m}$ is the coefficient of $b$ in a
    two-variable fit and is therefore a nonlinear function of the means, which
    amplifies the residual in windows where $|b|$ passes through zero. Six
    independent ensembles reproduce the plateau, so it is not an artifact of the
    nested prefixes used here. In all time-resolved panels the horizontal dotted line marks
    $\gamma_{m} = 0$. The two scans differ in how the ensembles are built:
    (a),(b) draw an independent ensemble of $M=200$ for each $\varepsilon$,
    while (c),(d) reuse nested prefixes of one $M=400$ ensemble, so successive
    points there share samples and are correlated by construction, which is why
    the $1/\sqrt{M}$ line in (d) is a guide rather than a fit. All four panels use the exact-derivative estimator, so
    what is plotted is $\gamma_{\rm exact} = \gA$ and no scheme component
    enters; the forward difference was computed alongside and agrees in sign at
    every $\varepsilon$, as quoted in the text.}
  \label{fig:robustII-eps}
\end{figure}

\begin{figure}[!tb]
  \centering
  \includegraphics[width=\columnwidth]{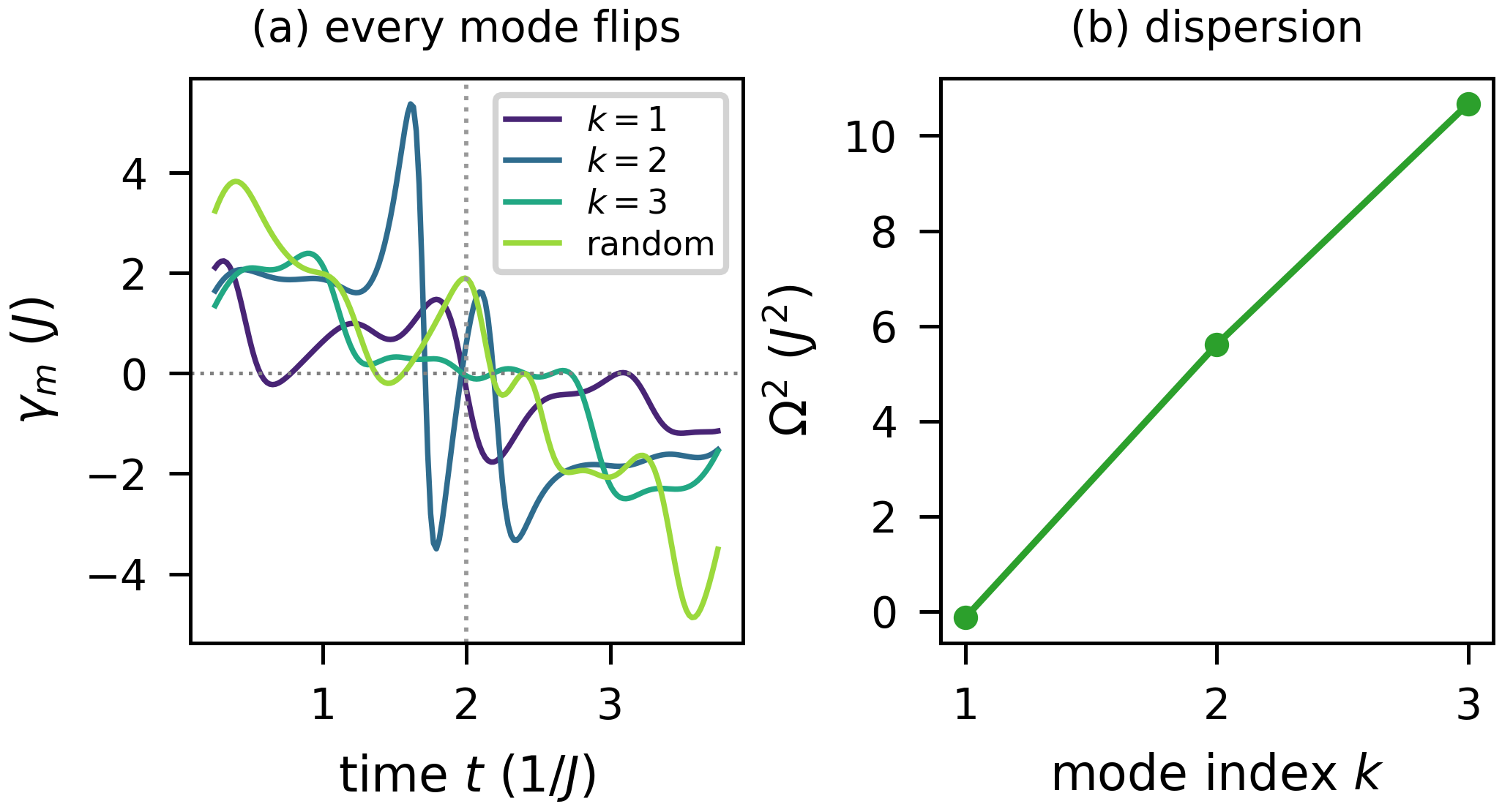}
  \caption{\textbf{Mode selection.} (a) Mode-layer friction for several seeded
    mode profiles, indexed by the wavenumber $k$ of Eq.~\eqref{eq:mode-ops},
    together with a random profile; every one reverses about the midpoint.
    (b) The fitted $\Omega^{2}$ (median over well-conditioned windows) versus
    mode index, which grows with $k$ as the dispersion of the chain requires;
    the softest mode sits at the resolution limit of the fit and comes out
    marginally negative. Both panels use the exact-derivative estimator.}
  \label{fig:robustII-modes}
\end{figure}

The reversal is not an artifact of a particular choice of conditioning
[Fig.~\ref{fig:robustII-eps}(a),(b)]. Varying
the boundary-modulation strength of Sec.~\ref{sec:resII-setting} over the whole
range we can access, $\varepsilon \in [0.15, 0.80]$, below which the conditioning is too weak to seed the mode and above which the ensemble becomes dominated by a few samples, the reversal survives at every $\varepsilon$, with
the exact and forward-difference estimators agreeing in sign throughout; the amplitude of $\gamma_{m}$ varies only between $0.54$
and $1.05$ over that range while the median overlap grows from $0.05$ to
$0.19$, so stronger conditioning increases the signal without changing the
phenomenon. The ensemble converges quickly [Fig.~\ref{fig:robustII-eps}(c),(d)]: the reversal is
already present at an ensemble size of $M = 50$ samples, and the reflection
residuals of the weighted means fall as $1/\sqrt{M}$
[Fig.~\ref{fig:robustII-eps}(d)], which is the precise sense in which the
protection is a property of the ensemble average rather than of
individual samples. The residual of the fitted friction saturates near $0.25$
beyond $M \simeq 100$; that floor is inherited from the nonlinearity of the
window fit and not from the ensemble, as the figure caption explains. It is also
independent of which collective coordinate is conditioned
[Fig.~\ref{fig:robustII-modes}]: the reversal occurs
for every seeded mode profile we tried, including a random one, and the fitted
$\Omega^{2}$ increases with the mode index, consistent with the dispersion of
the chain. Conditioning a mode that was not
seeded yields a coherent amplitude an order of magnitude smaller, confirming
that the boundary modulation conditions the intended mode.

\begin{figure}[!tb]
  \centering
  \includegraphics[width=\columnwidth]{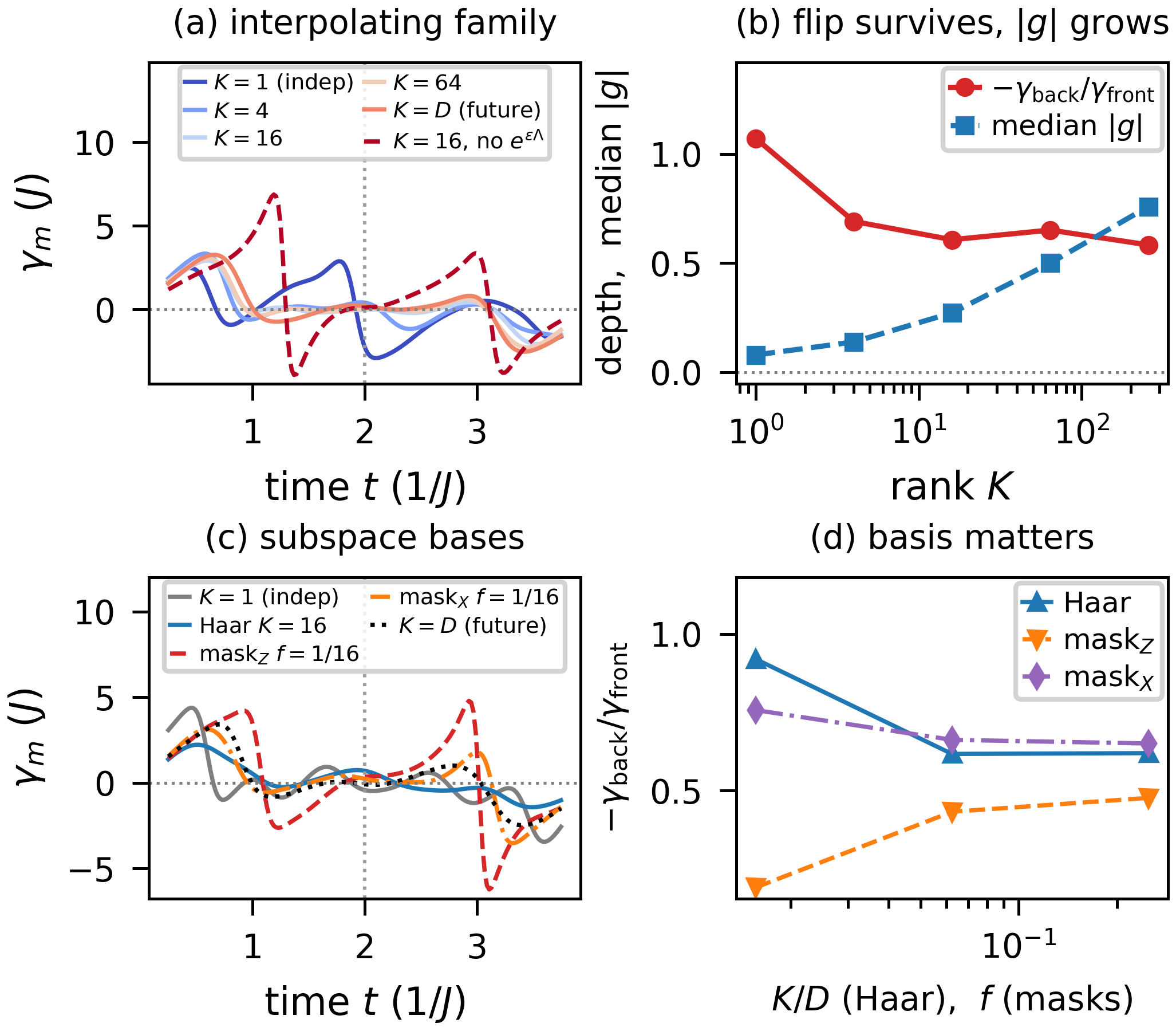}
  \caption{\textbf{Post-selection classes.}
    (a) The interpolating family $\varphi_{T} \propto
    e^{\varepsilon \Lambda}\Pi_{K}\ket{\psi(T)}$, labelled by the rank $K$ of
    the projector. Here and in (b), $\Pi_{K}$ projects onto a Haar-random
    $K$-dimensional subspace, drawn independently for each sample as the range
    of a $D \times K$ complex Gaussian matrix; this is the same construction as
    the Haar curves of (c) and (d), against which the basis-aligned masks are
    compared there. The two limits are the classes named elsewhere in the text:
    $K=1$ is independent conditioning and $K=D=256$ is future-consistent
    conditioning, so the family interpolates continuously between them. The
    dashed curve is the control that omits the boundary modulation,
    $\varphi_{T} \propto \Pi_{K}\ket{\psi(T)}$ at the same rank $K=16$; it is
    the only member that does not reverse, which is what identifies the
    modulation, and not the projection, as the cause of the reversal.
    (b) The reversal depth survives while the brightness
    $|g|$ grows monotonically with $K$. (c),(d) Subspace masks in the $Z$ and
    $X$ bases compared with Haar-random subspaces of equal rank fraction: the
    $X$-basis mask reproduces Haar behavior, the $Z$-basis mask does not.
    This is also why the two panels use different abscissae. Every member of the
    family in (b) is a Haar subspace, so its rank $K$ labels it unambiguously,
    whereas (d) compares constructions whose natural parameters differ and
    therefore needs the fraction they have in common: $K/D$ for the Haar points
    and $f$ for the mask points, which is what the axis label of (d) records.
    The masks are labelled by the retained fraction $f$ and the Haar subspaces by
    their rank $K$, because the two are specified differently: a Haar subspace
    has exactly rank $K$, whereas a mask keeps each basis state independently
    with probability $f$, so its rank is $fD$ only on average. At $D=256$ that
    distinction is not cosmetic, $f=1/16$ giving a rank of $16 \pm 4$ from
    sample to sample. Using $f$ also makes the same conditioning specifiable at
    both system sizes, since it does not scale with $D$.
    All panels use the exact-derivative estimator at the mode layer.}
  \label{fig:robustII-class}
\end{figure}

The dependence on the post-selection class isolates \emph{which} feature of
the future boundary causes the reversal. We construct a one-parameter family, $\varphi_{T} \propto
e^{\varepsilon \Lambda}\Pi_{K}\ket{\psi(T)}$, that interpolates exactly between
independent rank-one conditioning at $K=1$ and future-consistent conditioning at
$K = D$, $K$ being the rank of the projector $\Pi_{K}$, which projects onto a
Haar-random subspace redrawn for every sample. Every member reverses,
with the reversal depth, the ratio $-\gamma_{\rm back}/\gamma_{\rm front}$ of the second- to the first-half window average of $\gamma_{m}$, varying only
between
$0.58$ and $1.07$, while the overlap $|g|$ grows monotonically from $0.08$ to
$0.76$ [Fig.~\ref{fig:robustII-class}]. The control that identifies the
mechanism is the unmodulated projection $\Pi_{K}\ket{\psi(T)}$: it does
\emph{not} reverse. What causes the mode arrow to turn is therefore the
modulation imposed at the future boundary, not the subspace structure of the
post-selection. Consistently, the future-consistent class, which breaks the involution since $\varphi_{T}$ is then a function of $\psi$, still reverses
but loses the protection of antisymmetry, its midpoint defect being displaced
from zero while the independent class remains consistent with zero. The one combination
in which neither protection applies, namely the dictionary layer of a class that breaks the involution, is accordingly the one place where conditioning leaves
no trace of reversal at all.

\begin{figure}[!tb]
  \centering
  \includegraphics[width=\columnwidth]{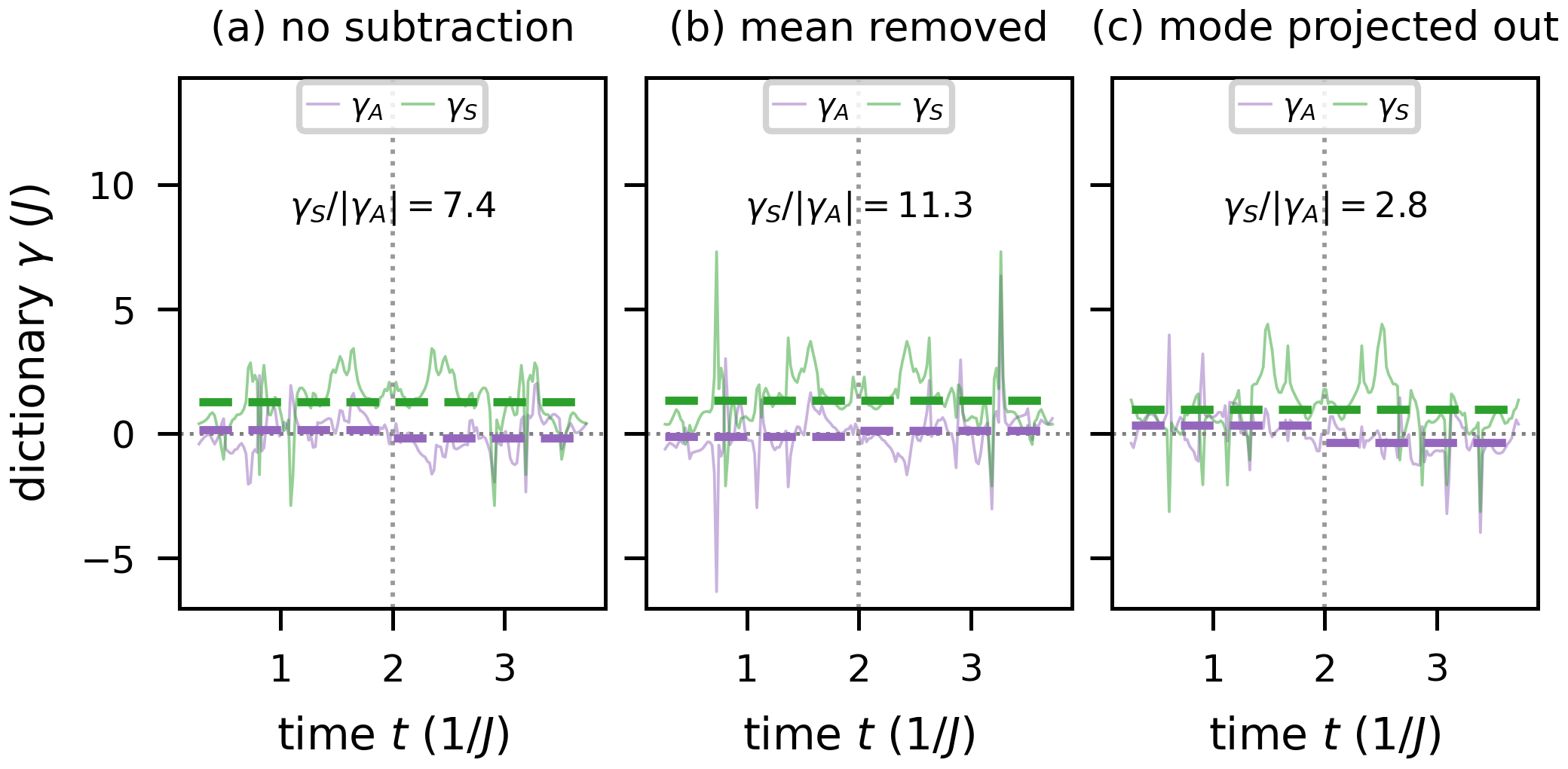}
  \caption{\textbf{The fluctuation-layer $\gA$ is a projection artifact.}
    Dictionary-layer friction for the three constructions of the text. All
    three use the same $29$-dimensional hydrodynamic dictionary and differ only
    in what is removed from the samples beforehand: (a) nothing,
    (b) the $|g|^{2}$-weighted ensemble mean, subtracted from every sample, and
    (c) the seeded mode direction,
    projected out of the $Z$ and current sectors. The inequality
    $\gS > |\gA|$ is a statement about the dashed levels, not about the traces:
    green is the window average of $\gS$, purple the two half-window averages of
    $\gA$, and the ratio of the former to the larger of the latter is printed in
    each panel. Read that way, the sign and size of $\gA$ depend on the
    construction, changing sign between (a) and (b), while the average $\gS$
    exceeds $|\gA|$ in all three by factors of $7.4$, $11.3$ and $2.8$, which is
    the inequality that forbids reversal. Pointwise the inequality holds in
    $69$--$82\%$ of windows; as everywhere in this paper, the claim is about
    the averages.
    Here $\gA$ and $\gS$ are the half-sum and half-difference of the forward and
    backward dictionary regressions at $\Delta t_{\rm cg} = C\,dt = 0.04$, the
    same resolution as Fig.~\ref{fig:bridge}(d); this figure
    is about the differencing-based decomposition, so no exact-derivative curve
    is shown.}
  \label{fig:modeproj}
\end{figure}

A second group of checks concerns the interpretation of the fluctuation-layer
$\gA$. We attempted to
attribute the small residual $\gA$ to contamination by the seeded coherent mode,
removing the mode in two independent ways: subtracting the $|g|^{2}$-weighted
ensemble mean from every sample, and projecting the mode direction out of the
$Z$ and current sectors of the dictionary. Neither removes it. Moreover, its
sign and magnitude depend on which removal is used
($\gA$ in the first half being $+0.17$, $-0.12$ and $+0.36$ for the three
constructions), so $\gA$ in the fluctuation layer is a property of the
dictionary projection and not of the ensemble [Fig.~\ref{fig:modeproj}]. We
therefore refrain from interpreting it. What the argument actually needs is
only the inequality $\gS > |\gA|$, equivalent to $\gfwd > 0$ throughout, and
that holds with a margin of $2.8$ to $11.3$ in every construction, the scheme component being stable at $\gS \simeq 1.0$--$1.4$ throughout.

\section{Results III: scaling to a million dimensions}
\label{sec:resIII}

\subsection{Size-independence of the involution}
\label{sec:resIII-pairtest}

The theorem of Sec.~\ref{sec:resII} is algebraic, and should therefore hold at
any size. We verify this directly at $N = 20$, i.e.\
$D = 2^{20} = 1\,048\,576$, using a matrix-free sparse implementation
(Appendix~\ref{app:numerics}). The test propagates a single reflection-paired
pair of samples from a fixed seed and compares the two trajectories element by
element. The sample-level identity
\eqref{eq:sample-identity} holds with
$|g'-g| = 2.6\times10^{-17}$, mode deviations of $10^{-14}$--$10^{-13}$, and
dictionary deviations of $2.0\times10^{-14}$ (time-even) and
$3.1\times10^{-14}$ (time-odd) across all $77$ dictionary elements, with
$\braket{\varphi(t)}{\psi(t)}$ constant to $1.3\times10^{-16}$. Because that
run takes hours at full length, we also report the same test over a shortened
conditioning window, $T = 0.4$ in place of $T = 4$, which runs in minutes on a
workstation and gives the same picture: $|g'-g| = 1.1\times10^{-16}$, mode
deviations of $4\times10^{-14}$--$4\times10^{-13}$, dictionary deviations of
$6.0\times10^{-14}$ and $8.6\times10^{-14}$, and
$\braket{\varphi(t)}{\psi(t)}$ constant to $5.5\times10^{-17}$. The identity is
a per-step algebraic property, so it does not improve or degrade with the
length of the window; the shortened test is the cheap way to reproduce it.

\begin{figure}[!tb]
  \centering
  \includegraphics[width=\columnwidth]{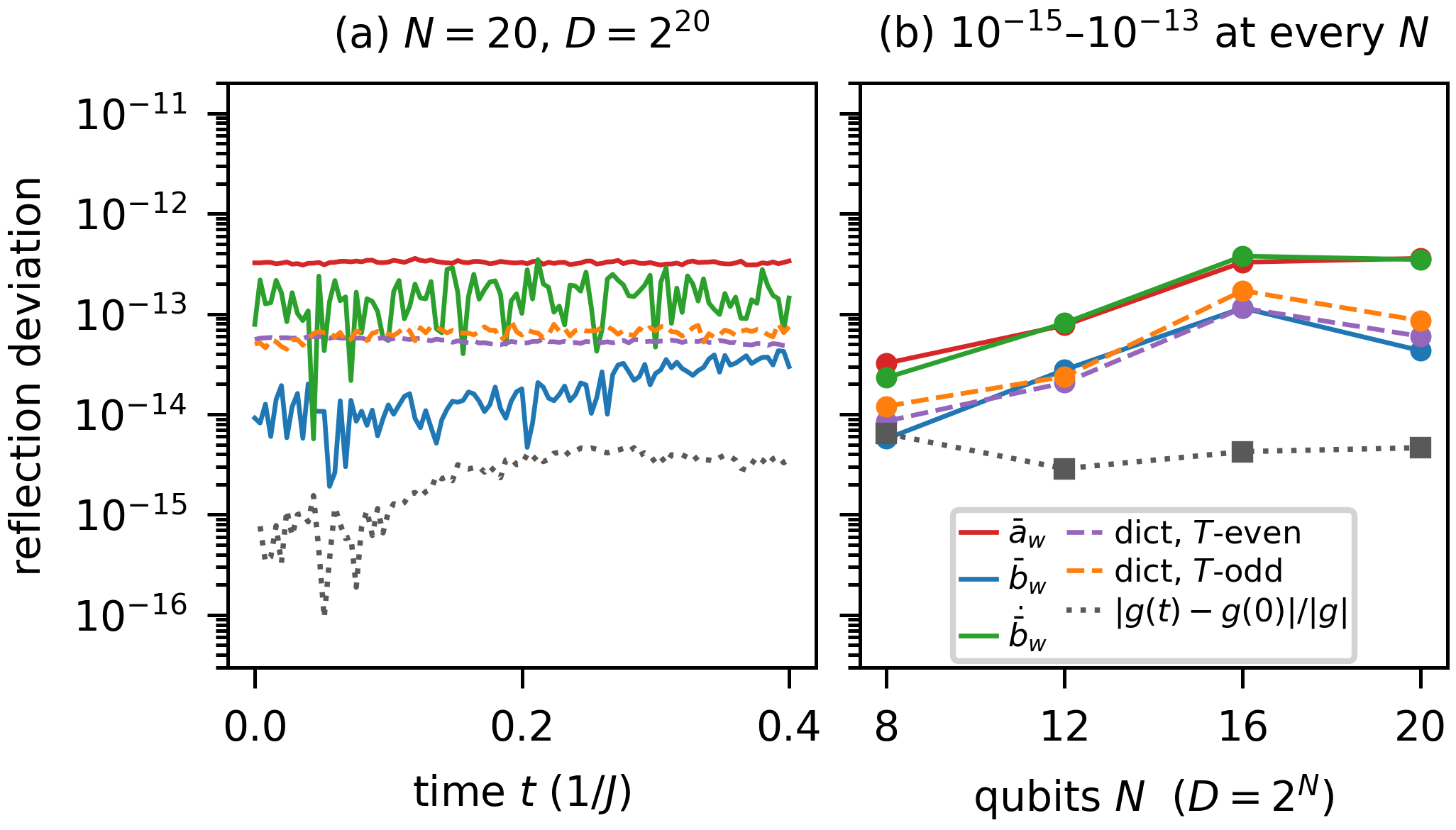}
  \caption{\textbf{The involution holds step by step, at every size.}
    Deviation from the sample-level identity~\eqref{eq:sample-identity} for a
    single reflection-paired pair of samples with independent conditioning,
    over the shortened window $T = 0.4$ ($n_{T} = 101$).
    (a) Resolved in time at $N = 20$: the mode observables
    $\bar a_{w}$, $\bar b_{w}$, $\dot{\bar b}_{w}$; the largest deviation over
    the time-even ($Z$, $ZZ$) and time-odd ($J$, $K$) dictionary elements; and
    the constancy of the overlap $g$. Nothing drifts with $t$, which is what
    distinguishes a per-step identity from one that holds only in the mean.
    (b) The same maxima against size. They stay between $6\times10^{-15}$ and
    $4\times10^{-13}$ over a factor of $4096$ in $D$, growing by a factor of
    $11$ rather than with the dimension, so the verification is not a
    small-$D$ accident. The full-length run at $N = 20$ quoted in the text
    gives the same levels.}
  \label{fig:pairtest}
\end{figure}

Figure~\ref{fig:pairtest} resolves that test in time and repeats it at four
sizes. The deviations do not drift with $t$ [Fig.~\ref{fig:pairtest}(a)], as
they should not for an identity that holds step by step, and they stay between
$6\times10^{-15}$ and $4\times10^{-13}$ from $N = 8$ to $N = 20$
[Fig.~\ref{fig:pairtest}(b)], that is, while $D$ grows by a factor of $4096$:
the largest deviation grows by a factor of $11$ over that range, from
$3.2\times10^{-14}$ to $3.6\times10^{-13}$, and $|g'-g|$ stays at $10^{-16}$
throughout. The residual therefore tracks the accumulation of round-off, not
the dimension, which is the numerical content of the theorem's being
algebraic.

The same run quantifies the obstruction that motivates
Sec.~\ref{sec:resIII-selfaverage}: for independent post-selection at $N=20$ the
overlap of the pair-test sample is only $|g| = 7.44\times10^{-3}$, a tenth of
its typical $N=8$ value (the ensemble median of the class,
Table~\ref{tab:q20}, is $3.5\times10^{-3}$).

\subsection{The two-layer structure at \texorpdfstring{$D=2^{20}$}{D=2^20}}
\label{sec:resIII-twolayer}

\begin{figure*}[!tb]
  \centering
  \includegraphics[width=\textwidth]{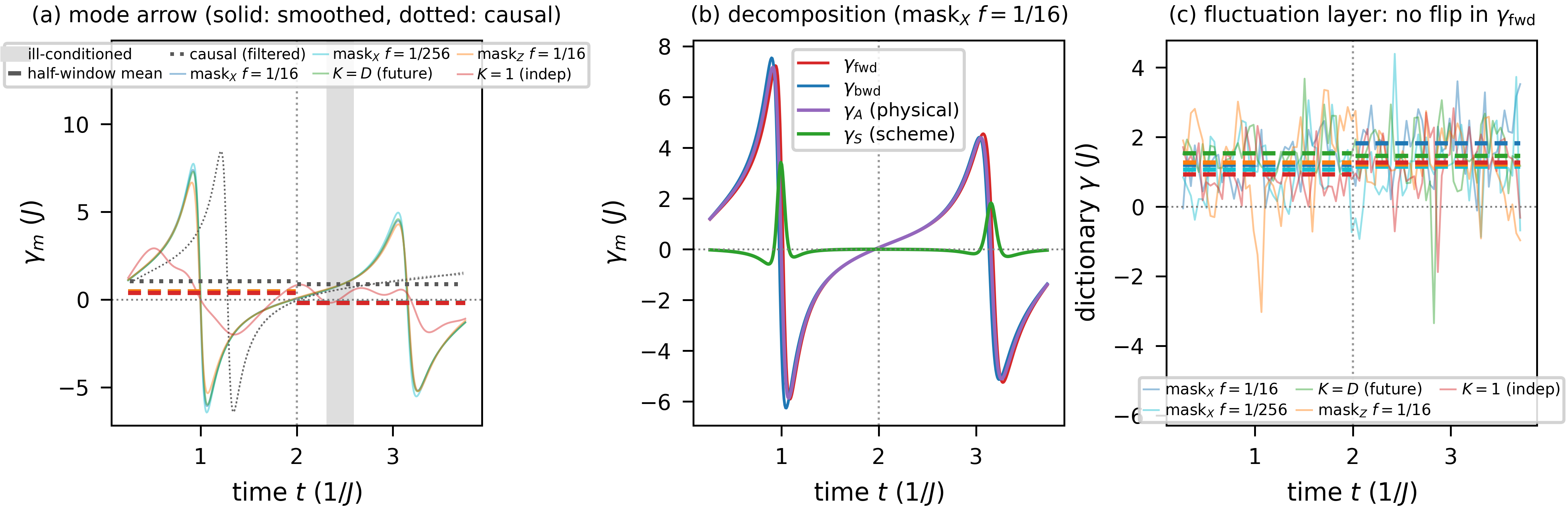}
  \caption{\textbf{Two-layer arrow of time at $D = 2^{20}$.}
    (a) Mode-layer friction from the exact derivative; colors label
    post-selection classes as in the legend. Thin solid curves are the
    conditioned $\gamma_{m}(t)$; heavy dashed levels in the same colors are
    their half-window averages over the well-conditioned windows, all five
    within $0.09$ of one another (Table~\ref{tab:q20}), so they overlap and only
    the last drawn, $K=1$, shows on top. The thin gray dotted curve is the
    causal (``filtered'') reference, the same fit applied to the expectation
    value conditioned on the past alone, and the heavy gray dotted levels are
    its half-window averages, $+1.04$ and $+0.87$; it is drawn in gray because
    it is class-independent, the five classes giving it to within $3\%$. At this
    size the oscillation amplitude exceeds the half-window mean by a factor of
    about $16$, so the reversal is read from the sign change of the heavy dashed
    levels and not from the traces, while the heavy dotted ones stay positive in
    both halves. The gray band marks windows excluded because the fit is
    ill-conditioned in them. The fit
    \eqref{eq:mode-fit} reads $\gamma_{m}$ off the coefficient of
    $\bar b_{w}$, so it loses its grip wherever the mode velocity is small: the
    band is the interval, here $t = 2.31$--$2.59$, in which
    $\|\bar b_{w}\|$ over the window drops below $15\%$ of its largest value,
    and $\gamma_{m}$ there is a ratio of two small numbers whose sign is set by
    noise rather than by the dynamics. Appendix~\ref{app:conventions} gives the
    convention and its consequences. The four classes built with the same boundary modulation
    collapse onto one another, while independent conditioning is offset.
    (b) Decomposition of the mode-layer friction into $\gA$ and $\gS$ for the
    mask$_X$ $f=1/16$ class, with the two one-sided estimators it is built from.
    (c) Fluctuation layer, forward difference, same two-line convention: faint
    solid traces are the raw window fits of $\gfwd$ on the $77$-dimensional
    dictionary and the heavy dashed levels their half-window averages, the
    dictionary-layer column of Table~\ref{tab:q20}. These do \emph{not} overlap,
    running from $+0.92$ to $+1.82$, and all are positive in both halves, which
    is the absence of reversal; the traces alone do not show it, swinging between
    $-3.4$ and $+4.4$. No windows are excluded here, unlike (a): the
    ill-conditioning of Appendix~\ref{app:conventions} is specific to the
    two-variable mode fit.
    Panels (b) and (c) use $\Delta t_{\rm cg} = 0.04$
    (Table~\ref{tab:params}); as at $N=8$, the exact derivative carries no
    scheme component and (a) is therefore free of it.}
  \label{fig:q20}
\end{figure*}

Table~\ref{tab:q20} collects the ensemble results for five post-selection
classes, $1016$ samples in total; Fig.~\ref{fig:q20} shows the underlying
friction curves. Every class reverses at the mode level, with
bootstrap confidence intervals excluding zero in both halves for the four
bright classes [Fig.~\ref{fig:q20}(a)]; no class reverses at the dictionary
level [Fig.~\ref{fig:q20}(c)]; and the causal
(``filtered'') reference, the same fit applied to the ordinary expectation value conditioned on the past alone, never reverses. The fluctuation-layer margin lies between
$|\gA|/\gS = 0.09$ and $0.27$ across the five classes (Table~\ref{tab:q20}),
to be compared with
$0.135$ at $N=8$: the inequality $\gS > |\gA|$ holds with a factor of
$3.7$--$11$ to spare at both sizes, across a factor $4096$ in Hilbert-space
dimension. (For the classes that break the involution the
decomposition~\eqref{eq:decomp} is applied operationally, $\gA$ and $\gS$
being the half-sum and half-difference of the two estimators; its antisymmetry
protection is exact only for independent conditioning,
cf.\ Sec.~\ref{sec:resII-robustness}.)

Because the half-window averages of $\gamma_{m}$ are small residuals of a
large oscillation at this size (the oscillation amplitude exceeds the first-half mean by a factor $\sim 16$), we also report the
convention-insensitive reflection residuals of
Appendix~\ref{app:conventions}. For every bright class the physical component
leans antisymmetric, $\rho_{A}/\rho_{S} = 0.35$--$0.37$ for $\gA$
[Fig.~\ref{fig:q20}(b)]; the independent class, whose weights concentrate on a
few samples, gives $0.53$. For $\gS$ both
residuals are small and comparable in the bright classes,
$\rho_{A} \simeq \rho_{S} \simeq 0.1$ (against $0.41$ and $0.22$ for the independent
class): at this
size the mode-layer $\gS$ is a small, noise-dominated correction, so neither signature dominates. This is consistent with a quantity that carries no arrow, though
weaker than the $\rho_{S} \ll \rho_{A}$ signature such a quantity would show at high
signal-to-noise.

\begin{table*}[t]
  \caption{\textbf{Two-layer structure at $N=20$ ($D = 2^{20}$).}
    For each post-selection class: number of samples $M$, median overlap
    $|g|$, mode-layer friction $\gamma_{\rm exact}$ (first/second half of the
    conditioning window, ill-conditioned windows excluded; see
    Appendix~\ref{app:conventions}), reversal depth
    $-\gamma_{\rm back}/\gamma_{\rm front}$, where $\gamma_{\rm front}$ and
    $\gamma_{\rm back}$ are those two half-window averages, the bootstrap
    reversal probability $P({\rm flip})$,
    dictionary-layer friction (first/second half), and the no-flip margin
    $|\gA|/\gS$ of the dictionary layer, defined as the larger of the two
    half-window averages of $|\gA|$ divided by the mean of $\gS$.
    mask$_{Z}$ denotes a
    computational-basis subset projector and mask$_{X}$ its Hadamard conjugate
    (Sec.~\ref{sec:disc-noncommut}), with $f$ the fraction of basis states
    retained: each is kept independently with probability $f$, so the rank of
    the projector is $fD$ in expectation rather than exactly. Depths are computed from unrounded
    values. The dictionary-layer columns, and with them the margin
    $|\gA|/\gS$, are evaluated at $\dtcg = 0.04$; $\gamma_{\rm exact}$ needs no
    such choice. The half-window averages, and hence the depths, are quoted at the
    exclusion threshold $q = 0.15$ of Appendix~\ref{app:conventions}; their
    sign is independent of that convention but their magnitude is not
    (Table~\ref{tab:qscan}), which is why the threshold-insensitive residuals
    are reported alongside them in the text. The causal reference is
    class-independent (the quoted pair is representative; the per-class values
    agree to a few percent), as it must be.}
  \label{tab:q20}
  \begin{ruledtabular}
  \begin{tabular}{lccccccc}
    class & $M$ & median $|g|$ & $\gamma_{\rm exact}$ front/back
          & $-\gamma_{\rm back}/\gamma_{\rm front}$
          & $P({\rm flip})$ & dict.\ front/back & $|\gA|/\gS$ \\
    \hline
    mask$_X$ $f=1/16$   & 320 & 0.472  & $+0.461 / -0.198$ & 0.429 & 1.00
                        & $+1.21 / +1.82$ & 0.20 \\
    mask$_X$ $f=1/256$  & 128 & 0.168  & $+0.458 / -0.181$ & 0.395 & 1.00
                        & $+1.06 / +1.15$ & 0.27 \\
    future-consistent   & 191 & 0.629  & $+0.458 / -0.191$ & 0.416 & 1.00
                        & $+1.53 / +1.46$ & 0.21 \\
    mask$_Z$ $f=1/16$   & 185 & 0.160  & $+0.468 / -0.201$ & 0.430 & 1.00
                        & $+1.27 / +1.21$ & 0.09 \\
    independent         & 192 & 0.0035 & $+0.379 / -0.189$ & 0.500 & 0.84
                        & $+0.92 / +1.27$ & 0.18 \\
    \hline
    causal (filtered)   & --- & ---    & $+1.04 / +0.87$   & ---   & ---
                        & --- & --- \\
  \end{tabular}
  \end{ruledtabular}
\end{table*}

\subsection{Self-averaging removes the brightness obstruction}
\label{sec:resIII-selfaverage}

The mode-layer result depends only weakly on the post-selection class:
$\gamma_{\rm exact}$ in the first half lies
between $+0.458$ and $+0.468$ for the four bright classes, whose overlaps
$|g|$ differ by a factor of $4$; the independent class, dimmer by a further
factor of $46$, gives $+0.38$ with a bootstrap interval wide enough to contain
the bright-class values. Comparing the mode trajectories themselves, the
future-consistent class and the two $X$-basis mask classes agree with one
another to a relative $5$--$8\times 10^{-4}$ in $L^{2}$ norm, the
$Z$-basis mask class to $5\times10^{-3}$, and the independent class to
$0.22$.

The explanation is that the $|g|^{2}$-weighted average
\eqref{eq:selfaverage}
depends, for post-selections drawn independently of the pre-selection, on the
post-selection ensemble only through its second moment
$\mathbb{E}[\ket{\varphi_{T}}\bra{\varphi_{T}}]$, which is set by the boundary
modulation $e^{\varepsilon \Lambda}$. Indeed the independent class agrees
with the others within its statistical precision. The four bright classes are
all constructed from $\ket{\psi(T)}$ and are therefore not covered by this
argument; that they collapse onto one another at the $10^{-3}$ level or below
is an empirical observation, indicating that at this size the $\psi$-dependent
structure of the post-selection self-averages away and only the shared
boundary modulation survives. The collapse is not uniform across the four:
the three classes whose mask acts in a basis conjugate to the dictionary agree
to $5$--$8\times10^{-4}$, while the $Z$-basis mask, which commutes with the
$Z$ sector of the dictionary and therefore conditions it weakly
(Sec.~\ref{sec:disc-noncommut}), sits an order of magnitude off at
$5\times10^{-3}$; it remains two orders below the spread of the independent
class. At $D = 256$
self-averaging is weak and the class details (rank fraction, and the basis in which a subspace mask is applied) do survive, which is what our $N=8$ scans
detect. The two observations are therefore consistent, the class dependence
being a small-$D$ phenomenon.

This resolves the brightness problem. Since
the bright classes agree at the $10^{-3}$ level or below and the independent class is
consistent within its precision, one may use the brightest
available class: $|g| = 0.47$ for a Hadamard-conjugated subspace mask, against
$3.5\times10^{-3}$ for independent conditioning. Independent conditioning,
although it is the class for which the involution holds exactly, is not a
usable estimator at $N=20$: its reversal probability reaches only $0.84$ at
$M=192$, because the strongly fluctuating $|g|^{2}$ weights concentrate on a
few samples and the effective sample size is far below $M$. Its point estimate
is consistent with the other classes within its bootstrap interval, so this is
a variance problem, not a bias: the
involution makes the statement
exact, and the bright classes make it measurable.

\section{Discussion}
\label{sec:discussion}

\subsection{What conditioning can and cannot change}
\label{sec:disc-cannot}

Read together, the three results describe a single structure: what
conditioning on the future can change is constrained by symmetries. We list
the two sides explicitly. Conditioning
\emph{can} assign a property to an arm the particle does not occupy, give that
property its own conserved current, and reverse the friction of a coherent
hydrodynamic mode. It \emph{cannot} move the particle out of arm~U while
$\sigma_{z}$ is conserved there, cannot reverse the fluctuation-level friction
while $\gS$ dominates $\gA$, and cannot alter the causal record at first order
in the perturbation that breaks the protecting symmetry.

Every entry on the second side is protected by a
symmetry. The
particle's assignment is protected by a conservation law: $\wv{\Pi_{D}}$
vanishes by spin orthogonality, so no amount of tuning is required and no
perturbation that commutes with $\sigma_{z}$ can break it. This is why a junction that backscatters more than half the incident probability leaves the separation
intact three orders below the signal, and why the leakage does not scale with
the loss. The fluctuation-level arrow is protected by the reflection involution
through the inequality $\gS > |\gA|$. And the causal record is protected
parametrically: tilting the field breaks the first protection at order
$\sin\alpha$ in the conditioned description but only at order $\sin^{2}\alpha$
in $|g|$, so the conditioned description is one order more sensitive than the
statistics an experimenter collects.

This suggests a reading of the two-state formalism that does not require
deciding whether its assignments are ``real''. The formalism enlarges what
an observer may assign, and the symmetries of the underlying dynamics then
determine which of those assignments are robust. Robustness, not reality, is
the discriminating question, and it is answerable by measurement.

\subsection{Relation to smoothed quantum states and to fluctuation paths}
\label{sec:disc-smoothed}

The filtered/smoothed distinction we exploit is the one developed in the theory
of past quantum states and quantum
smoothing~\cite{Gammelmark2013, Tsang2009, Chantasri2021}, and the weak-value
continuity equation we use is
established~\cite{Wiseman2007, Kocsis2011}. What is new here is not the
smoothed description but what we extract from it: \emph{effective generators
and transport coefficients}, and with them a decomposition of the measured
friction into a boundary-condition component and an inference component. Where
the smoothing literature asks what an observer may estimate at an intermediate
time, we ask what effective \emph{dynamics} that observer would infer, and we
find that the answer splits into a part fixed by the boundary conditions and a
part fixed by the estimator.

That split gives an operational criterion for statements about opposing arrows
of time. Thermalization in two opposing time directions has been derived for
open systems from a time-symmetric formulation of the Markov
approximation~\cite{Guff2025}; our result is complementary and makes the
question quantitative. A reversal is visible at a given level of description
precisely when $|\gA| > \gS$ at that level and at the resolution at which the
generator is inferred, and $\gA$ and $\gS$ are separately
measurable. Whether an arrow reverses is therefore not a matter of
interpretation but of an inequality between two measurable quantities, and the
inequality can go either way depending on whether one looks at coherent modes
or at the fluctuation layer, and on how finely one differences. Only $\gA$
survives the exact-derivative limit, which is the sense in which it, and not
$\gfwd$, is the boundary-condition part.

The reflection involution of Sec.~\ref{sec:resII-involution} also has a
classical ancestor. In the Onsager--Machlup theory of fluctuation
paths~\cite{OnsagerMachlup1953} the functional whose minimum selects the most
probable trajectory between two prescribed boundary conditions is invariant
under time reversal, so reversal carries that trajectory to the most probable
trajectory with the two boundary conditions exchanged, and it is the
dissipative part of the dynamics that distinguishes the two directions of
traversal. Equation~\eqref{eq:involution} is the quantum,
pre-/post-selected version of that statement, and
Eq.~\eqref{eq:decomp} identifies which part of a measured friction inherits the
symmetry and which part violates it. The novelty on the quantum side is that
the violating part turns out to be an artifact of the observer's estimator
rather than a property of the ensemble.

\subsection{Relation to the Cheshire-cat debate}
\label{sec:disc-cheshire}

We make no ontological claim, and our results do not require one. What we
establish is operational and orthogonal to the dispute: the
separation is a dynamical structure of the time-symmetric description, absent
from the causal one [Fig.~\ref{fig:interferometer}(a)--(d)]; it is carried by a
pair of conservation laws rather than by a single instant; and it responds to
local perturbations in a way that identifies where the property is
[Fig.~\ref{fig:interferometer}(e),(f)]. None of these statements presupposes
that the property is ``disembodied''.

Indeed our findings are compatible with the deflationary readings. That the
effect can be understood as interference between pointer
states~\cite{Correa2015}, or as a feature of a contextual set of mutually
incompatible properties~\cite{Hance2023contextuality}, is consistent with
everything reported here: our weak polarization vector is null,
$\wv{\sigma_{x}}^{2} + \wv{\sigma_{y}}^{2} = 0$, which is precisely the kind of
structure one expects when a single complex amplitude is being read through two
incompatible projections. We quantify the structure; we do not adjudicate its
interpretation.

Where we do contribute to the debate is on detectability. Hance, Ladyman and
Rarity argued that the dynamical Cheshire cat is not detectable in the original
scenario, requiring a perturbation that admixes an orthogonal
component~\cite{Hance2024detectable}. Our response experiment is a perturbation
of this kind, implemented as a local field rather than as an admixture to the
initial state, and it makes the cost explicit and quantitative: the conditioned
response to a symmetry-breaking tilt is first order in $\sin\alpha$ while the
leakage into the post-selection statistics is only second order. An experiment
must therefore resolve a weak-measurement signal one order in the perturbation
before the effect shows up in its counting rates. This quantifies the difficulty and identifies
the figure of merit an experiment would have to reach.

\subsection{What makes a post-selection informative}
\label{sec:disc-noncommut}

Our subspace scans point to a general criterion for the \emph{strength} of a
conditioning, which we state here because it is useful independently of the
present application: a post-selection subspace conditions an observable weakly
to the extent that it commutes with that observable.

The extreme case is instructive. Projecting onto an eigenspace of a conserved
charge carries no information about the future whatsoever: since the projector
commutes with the evolution, $\Pi_{Q}\ket{\psi(T)}$ is the same sector
component of $\ket{\psi(0)}$, and the ``post-selection'' is a statement about
the initial state. This is not a small correction: the conditioning fails
entirely, which disqualifies the total magnetization,
otherwise the natural choice in a spin chain, as a conditioning observable.
Basis-subset masks in the $Z$ basis sit between the extremes: being diagonal,
they commute with the $Z$ sector of our dictionary, and they condition the mode
measurably less than Haar-random subspaces of the same rank fraction. Conjugate
the same masks by a Hadamard transform, so that they no longer commute with the
$Z$ sector, and Haar behavior is recovered. A mask is built by retaining each
basis state independently with probability $f$, which is why we label it by
that fraction rather than by a rank: unlike a Haar subspace, which is
constructed with exactly $K$ dimensions, a mask has rank $fD$ only on average.
The fraction is also the quantity that transfers between system sizes, since
the same $f$ specifies the same conditioning at $D = 256$ and at
$D = 2^{20}$, where building an explicit rank-$K$ Haar subspace is not
feasible in the first place.

At $D = 2^{20}$ this distinction disappears, for a reason that inverts the
usual expectation. For independently drawn
post-selections the $|g|^{2}$-weighted average
\eqref{eq:selfaverage} depends on the post-selection ensemble only through its
second moment, and empirically the same collapse occurs for the classes built
from the evolved state: once $D$ is large enough for self-averaging, classes
sharing a boundary modulation become indistinguishable, and the
class dependence mapped at $D = 256$ is a small-system
phenomenon. Large systems are in this respect \emph{simpler} than small ones:
the design freedom that matters shifts from the structure of the post-selected
subspace to the second moment, that is, to the boundary modulation itself.

\subsection{Limitations}
\label{sec:disc-limits}

Five limitations should be noted. First, the interferometer of
Sec.~\ref{sec:resI} is a single-particle system while the chain of
Secs.~\ref{sec:resII}--\ref{sec:resIII} is many-body: what unifies the two
halves of this paper is the method, not the model. This is intentional: the Cheshire structure requires a path degree of freedom and an
internal one, transport requires a thermalizing many-body system, and the point
is that one tool addresses both. It does mean, however, that no single system exhibits all of the phenomena we report.

Second, the exact statements of Sec.~\ref{sec:resII} require a real symmetric
Hamiltonian and a conjugation-invariant measure. Both hold for the XXZ family
studied here, but a complex Hamiltonian, a Peierls flux for instance, breaks the involution as we have written it. We expect the correct generalization to
compose $R$ with a time-reversal operation, and the decomposition to survive
with $\gA$ redefined accordingly; we have not carried this out.

Third, the ensemble-level symmetry at $N = 20$ is verified at the level of the
per-sample identity \eqref{eq:sample-identity}, not of the finite-$M$ ensemble.
The exact statement is one about the ensemble as a whole, and recovering it at finite $M$
requires the reflection-paired sampling we use at $N = 8$; we have implemented
and validated the paired sampler at $N = 20$ but have not run the corresponding
campaign, which is why Table~\ref{tab:q20} reports bootstrap significance
rather than machine-precision antisymmetry.

Fourth, the fluctuation-layer $\gA$ should not be given a physical reading. Its
sign and magnitude depend on how the dictionary is constructed
[Fig.~\ref{fig:modeproj}], so it is a projection artifact; only the inequality
$\gS > |\gA|$, and hence the absence of reversal, is construction-independent.
We have been careful to phrase the fluctuation-layer claim as the inequality
alone.

Fifth, that inequality is not resolution-independent. $\gS$ is what the
differencing injects and grows with $\Delta t_{\rm cg}$, so the margin at the
fluctuation layer is a property of the pair (ensemble, estimator) and shrinks as
the estimator is refined, until in the exact-derivative limit $\gS$ vanishes and
the layer reverses like the mode layer (Sec.~\ref{sec:resII-twolayer} and
Appendix~\ref{app:conventions}). The claim we make is therefore that at a stated
inference resolution the two layers behave oppositely, and that the difference
between them, a factor of $34$ in $\gS$ at fixed $\Delta t_{\rm cg}$, is large
enough that no reasonable choice makes the mode layer fail to reverse. What is
resolution-independent is the decomposition itself and the antisymmetry of
$\gA$. We add that the exact-derivative comparison at the fluctuation layer was
carried out at $N=8$, where the second derivatives of all $29$ dictionary
elements are affordable; at $N=20$ we computed the exact derivative for the mode
observables only, so the statement that both layers reverse under the exact
estimator rests on the theorem, which is size-independent, plus the $N=8$
measurement, not on a direct $N=20$ measurement of the dictionary layer.

\section{Conclusion}
\label{sec:conclusion}

We have introduced two-state gEDMD, a data-driven extraction of effective
generators from pre- and post-selected quantum trajectories, and used it to
separate what an observer's conditioning can change from what it cannot. The
enabling observation is that weak values satisfy
$dA_{w}/dt = i\langle[H,A]\rangle_{w}$ exactly: this
supplies an exact-derivative baseline, and with it the ability to distinguish a
transport coefficient from an artifact of the estimator that measured it. On
that basis a single reflection involution forces every measured friction to
split uniquely into an antisymmetric component carrying the boundary-condition
physics
and a symmetric component generated by the differencing scheme, and the two
are separately measurable as $(\gfwd \pm \gbwd)/2$. At a fixed inference
resolution the arrow of time
accordingly has two layers: coherent modes reverse at the midpoint of the
conditioning window, while the fluctuation layer does not, because the scheme
component it receives is $34$ times larger than at the mode layer and buries
the reversal.
In the exact-derivative limit the scheme component vanishes and both layers
reverse, so the second layer's immunity is a property of the description rather
than of the ensemble. Applied to a port-defined interferometer, the
same framework renders the quantum Cheshire cat a statement about property currents, with the particle and its polarization obeying separate conservation laws, and with a local field rotating only the property's phase, at exactly twice
the field strength, and appearing in the extracted generator as a rigid
imaginary spectral shift while the particle's weak density stays invariant to
machine precision.

Two features of the result extend beyond the specific models. The
first is that the structure is stable under scaling: it persists from $2^{8}$ to
$2^{20}$ Hilbert-space dimensions with the fluctuation layer's no-flip
inequality holding with a wide margin at both sizes, and at large dimension
self-averaging makes
the answer insensitive to the post-selection class. This removes the
$|\braket{\varphi}{\psi}| \sim 2^{-N/2}$ obstruction and means that the
brightest, most nearly realizable conditioning may be used without changing the
physics. The second is methodological. Whether a conditioned observer sees an
arrow of time is not a matter of interpretation but of an inequality between two
quantities, $|\gA|$ and $\gS$, both of which that observer can measure from the
same data. This part of the framework is portable: wherever
effective dynamics is inferred from trajectories under boundary conditions at both ends, whether in quantum trajectory experiments, in smoothed estimation, or in classical fluctuation-path settings, the same decomposition should apply, and
the same question should be asked of any reported friction before it is
interpreted as physics.

\section*{Code availability}
\label{sec:code}

All code required to reproduce the results of this paper will be made
available upon publication at
\url{https://github.com/saitos-lab/two-state-gedmd} under the MIT license.
The repository contains the interferometer and many-body engines, the
robustness scans, and the analysis pipeline. Every number quoted here is
printed by one of those scripts as a labelled check, so the verification logs
behind the paper are regenerated by rerunning them; the raw trajectory data
($\sim 2$~GB) is regenerated the same way. The full-length single-sample
$N=20$ check of Sec.~\ref{sec:resIII-pairtest} and the $1016$-sample
campaign were run on the cluster acknowledged below; the shortened checks that
Fig.~\ref{fig:pairtest} is built from run on a workstation in minutes.

\section*{Data availability}
\label{sec:data}

No experimental data were generated. All results are produced by numerical
simulation, and the raw trajectory data ($\sim 2$~GB) are not deposited because
they are regenerated deterministically by the released code from the parameters
and random seeds specified in Appendix~\ref{app:numerics}; they are available
from the author on reasonable request. The figures are rebuilt from the
intermediate arrays that the same scripts write.

\section*{Author contributions}
\label{sec:contrib}

S.S. is the sole author: he conceived the study, proved the involution theorem,
wrote the simulation and analysis code, performed and verified all calculations,
produced the figures, and wrote the manuscript. The generative AI models named
in the Acknowledgments were used as tools under his direction, for discussion of
the research concepts, assistance with the numerical analysis and the
preparation of the figures, and structural and language editing; they are not
authors, and the author takes full responsibility for the content of the paper,
including all results and claims.

\begin{acknowledgments}
The author gratefully acknowledges the computational resources provided by the
NIFS Plasma Simulator (Project ID: NIFS25KIST066), where the large-scale
$N = 20$ campaign of Sec.~\ref{sec:resIII} was performed.
Additionally, the author acknowledges the use of the generative AI models
Gemini (Google) and Claude Code (Anthropic) for facilitating discussions on the
research concepts, assisting with the numerical analysis and the preparation of
the figures, and providing structural organization and English language editing
of the manuscript.
\end{acknowledgments}

\appendix

\section{Numerical implementation}
\label{app:numerics}

Table~\ref{tab:params} collects the parameters of the three settings. We set
$\hbar = 1$ throughout, so that couplings and fields are rates, times are
measured in units of the inverse coupling, and phases, such as the grin phase of Fig.~\ref{fig:interferometer}(e) which advances at $2B$, are in radians. Accordingly the extracted frictions $\gamma$ and precession rates are
quoted in units of $J$, the fitted $\Omega^{2}$ in units of $J^{2}$, and times
in units of $1/J$; overlaps, densities, residual ratios and reversal depths are
dimensionless. Unless stated otherwise, a quoted residual is the maximum
absolute deviation over all sites and recorded times.

\begin{table}[b]
  \caption{\textbf{Parameters of the three settings.} $\dtcg = C\,dt$ is the
    coarse-graining interval of the finite-difference estimators, $\tau_{w}$ the
    sliding-window width and $s$ the window stride, both in time steps
    ($\tau_{w}$ also given in time units in parentheses; note that the two
    chains use nearly the same physical window). The window grids satisfy
    the reflection closure discussed in Appendix~\ref{app:conventions}. The
    interferometer dictionary comprises six spatial moments per sector
    (particle and polarization). For the interferometer, $C$ applies only to the
    forward-difference row of Table~\ref{tab:scheme}; every other quantity in
    Sec.~\ref{sec:resI}, including the spectra of Fig.~\ref{fig:spectra}, uses
    the exact derivative and involves no differencing interval. The
    interferometer has no entry for $\tau_{w}$ or $s$ because it is not
    analyzed with a sliding window: its generator is regressed once, over the
    single interval $t \in [6.2, 11.8]$ inside the field window, whereas the
    chains are analyzed on windows of width $\tau_{w}$ advanced by $s$ steps.
    $M$ is the number of pre-/post-selected
    samples averaged as in Eq.~\eqref{eq:selfaverage}: the interferometer is a
    single deterministic trajectory and needs none, the $N=8$ ensemble is
    reflection-paired, and the $N=20$ entry is the per-class range
    ($1016$ samples in total, Table~\ref{tab:q20}). The robustness scans of
    Sec.~\ref{sec:resII-robustness} use $M = 200$ per point, and up to $400$
    where $M$ itself is varied.}
  \label{tab:params}
  \begin{ruledtabular}
  \begin{tabular}{lccc}
     & interferometer & XXZ $N=8$ & XXZ $N=20$ \\
    \hline
    $\dim\mathcal{H}$      & $160$   & $256$   & $1\,048\,576$ \\
    couplings              & $J=1$   & \multicolumn{2}{c}{$J=\Delta=1$,
                                                          $J_{2}=0.5$} \\
    $dt$                   & $0.01$  & $0.002$ & $0.004$ \\
    steps                  & $1601$  & $2001$  & $1001$ \\
    $T$                    & $16.0$  & $4.0$   & $4.0$ \\
    $C$ ($\dtcg$)          & $40$ ($0.4$) & $20$ ($0.04$) & $10$ ($0.04$) \\
    $\tau_{w}$ (steps)     & ---     & $251$ ($0.502$) & $125$ ($0.500$) \\
    stride $s$             & ---     & $10$    & $2$ \\
    modulation $\varepsilon$ & ---   & $0.35$  & $0.35$ \\
    ensemble size $M$      & 1  & $160$   & $128$--$320$ \\
    dictionary size        & $6$/sector & $29$ & $77$ \\
  \end{tabular}
  \end{ruledtabular}
\end{table}

\emph{Interferometer.} The ladder carries $m = 40$ sites per rail (extended to
$m = 64$ for the localized-junction scan of Fig.~\ref{fig:yjunction}, so that
the backscattered branch does not return from the input edge within the run)
and the Hamiltonian is Eq.~\eqref{eq:H-interf}. All elements
are rectangular pulses, so $H(t)$ is piecewise constant and the time-ordered
propagator is an \emph{exact} product of nine segment exponentials: the
trajectory carries no time-discretization error, and $dt$ merely sets the
recording interval. Each pulse has width $\tau = 0.4$; the beam splitters are
switched on at $t = 3.0$ and $t = 13.0$ with
$\int\!\lambda\,dt = \pi/4$ (a 50/50 split), and the spin rotator at
$t = 4.4$ with $\int\!\chi\,dt = \pi/2$ (which takes rail D from
$\ket{\uparrow}$ to $\ket{\downarrow}$). The uniform field of
Sec.~\ref{sec:resI-response} has $\Pi_{F} = \Pi_{D}$ or $\Pi_{U}$, magnitude
$B = 0.15$ and window $t \in [6,12]$; the coil scan instead uses
$B = 0.05$ on five consecutive sites, centered at sites $19$, $22$, $25$
and $28$, on throughout the arm interval. Unitarity holds by construction (each
segment
propagator is the exponential of a Hermitian matrix), and
$\braket{\varphi}{\psi}$ is constant to $7\times10^{-15}$ over the full run. The wave packet
has central wavenumber $q = \pi/2$, hence group velocity $2J\sin q = 2J$ and
minimal dispersion; it is launched at site $x_{0}=6$ with width $1.3$. The
launch site is set by two requirements. The packet has a finite width, so it
must start far enough from the near edge not to be truncated there: at
$x_{0}=6$ the edge is $4.6$ widths away and carries a density of
$2\times10^{-5}$, whereas launching at the first site would leave $0.8$ widths
and cut the packet in half. It must also stay away from both edges while the
conditioned quantities are being read, and with $v_{g}=2J$ it occupies sites
$16$ to $32$ during the analysis window between R and BS2, at least five widths
from either end, so that no reflection contaminates the arm region. Only at
$t=T$, where the post-selection is defined, does the freely propagated packet
approach the far edge, being centered near site $36$ of rail D; no conditioned
quantity is evaluated there.

\emph{Many-body engine.} The chain Hamiltonian is Eq.~\eqref{eq:H-xxz} at both
sizes. For $N = 20$ it is stored as a sparse
hopping matrix ($1.0\times10^{7}$ nonzeros) plus a diagonal vector, and
propagation uses a truncated Taylor series with $\|H\|\,dt \simeq 0.2$. All
observables are evaluated matrix-free: diagonal densities as batched inner
products against a stacked coefficient matrix, bond currents by index
permutation with $\pm 1$ sign vectors, and the second derivative $i[H,b]$ from
$H\ket{\varphi}$ and $H\ket{\psi}$ without ever forming an operator product. One
sample requires $85$--$105$ min on one core and $1.2$~GB, so ensembles are run
as independent worker processes. The post-selection classes are built at
$t=T$ as follows: independent conditioning uses a fresh seed,
Eq.~\eqref{eq:seeds}; future-consistent conditioning uses
$e^{\varepsilon\Lambda}\ket{\psi(T)}$; a Haar rank-$K$ subspace is the range of a
$D\times K$ complex Gaussian matrix, orthonormalized by QR and redrawn per
sample, which is affordable only at $N=8$; and a basis-subset mask keeps each
basis state independently with probability $f$, in the computational basis for
mask$_{Z}$ and in its Hadamard conjugate for mask$_{X}$, the latter applied with
a fast Walsh-Hadamard transform so that no $D\times D$ matrix is ever formed. \emph{Reading the friction off the dictionary layer.} The dictionary is ordered
in four blocks, the site densities $Z_{i}$, the bond currents
$J_{b}$, the correlators $Z_{i}Z_{i+1}$ and the symmetric bond operators
$K_{b}$, so that its size is $N + 3(N-1)$, namely $29$ at $N=8$ and $77$ at
$N=20$. The window regression of Sec.~\ref{sec:method-gedmd} returns the
generator on all four. We then retain the hydrodynamic pair
$\{Z_{i}, J_{b}\}$, of dimension $N + (N-1)$, and eliminate the remaining two
blocks by a Schur complement,
$L_{\rm cl} = L_{11} - L_{12} L_{22}^{+} L_{21}$, with subscript $1$ the
retained pair and $2$ the eliminated remainder, the pseudoinverse being taken
with a relative cutoff of $10^{-3}$. The dictionary-layer friction is the median
of $-\mathrm{Re}\,L_{\rm cl}$ over the diagonal current entries, excluding the
two end bonds, which the open spatial boundary contaminates. The median rather
than the mean is used because a single poorly conditioned bond would otherwise
dominate. This closure is the one referred to in Appendix~\ref{app:proof},
where it is shown to commute with the reflection signature and hence to preserve
Eq.~\eqref{eq:equivariance}. The implementation reproduces a dense reference
at $N=8$ to $10^{-13}$ in every observable; at $N=20$ the recorded mode
velocity $b$ agrees with the numerical derivative of $a$ to $5\times10^{-6}$,
and the exact second derivative $i[H,b]_{w}$ with the numerical derivative of
$b$ to $2\times10^{-5}$.

\section{Analysis conventions}
\label{app:conventions}

Three conventions are needed to make the statements of Sec.~\ref{sec:resII}
testable, and two of them change conclusions if omitted. We record all three.

\emph{Reflection closure of the window grid.} The equivariance relations
\eqref{eq:equivariance} relate a quantity at $t$ to the same quantity at $T-t$,
so they can only be tested at machine precision if the set of window centers is
itself invariant under $t \leftrightarrow T-t$. We therefore choose the number of
time steps, the window width and the stride so that the grid closes under
reflection, which makes array reversal an exact mirror operation: at $N=8$,
$n_{T}=2001$ with $\tau_{w}=251$ and $s=10$; at $N=20$, $n_{T}=1001$ with
$\tau_{w}=125$, $C=10$ and $s=2$ (Table~\ref{tab:params}). With an even
window width, or a stride that does not divide the reflected offset, the same
data yield a spurious residual of order $10^{-2}$ in
Eq.~\eqref{eq:equivariance}, which is easily mistaken for a physical breaking of
the symmetry.

\emph{Exclusion of ill-conditioned windows.} The two-variable fit
\eqref{eq:mode-fit} determines $\gamma_{m}$ from the
coefficient of $b$, and therefore degenerates when the mode velocity passes
through zero inside a window: $|\gamma_{m}|$ diverges there, with a sign set by
noise. We exclude windows whose $\|b\|$ falls below a fraction $q$ of its
maximum, taking $q = 0.15$ wherever the exclusion is applied, and
report the number of well-conditioned windows alongside every fit. At $N = 8$
this is a cosmetic choice, changing the half-window averages by $7$--$8\%$
($\pm 0.960 \to \pm 0.894$ for the exact estimator) and no conclusion, so the
$N=8$ numbers, figures and Table~\ref{tab:cgscan} are quoted on the full window
grid. At
$N = 20$ it is not: the oscillation amplitude of $\gamma_{m}$ exceeds its
half-window mean by a factor $\sim 16$, so the mean is a small residual of a
large oscillation. For the flagship mask$_{X}$ $f=1/16$ class, $36$ of the
$439$ windows are excluded at $q=0.15$ ($36$ also for mask$_{X}$ $f=1/256$
and for the future-consistent class, $41$ for mask$_{Z}$ $f=1/16$, and none
for the independent class, whose mode amplitude never comes close to zero),
and those $36$ windows are enough to move the second-half average from
$-0.030$ to $-0.198$.

The threshold therefore has to be reported with the number. Table~\ref{tab:qscan}
gives the scan. Two features of it matter. First, the sign of the second-half
average, and hence the reversal itself, is the same for every $q$ we examined,
including $q=0$; what depends on the convention is the depth, which varies by a
factor of $20$ over the range. The depths in Table~\ref{tab:q20} are therefore
convention-dependent summaries, not measurements of a converged quantity, and
there is no plateau in $q$ from which a preferred value could be read off.
Second, the smallest window norm at $N=20$ is $0.105$--$0.110$ of the maximum
for the bright classes ($0.162$ for the independent class), so nothing at all is
excluded below $q \simeq 0.11$ and the value $q = 0.10$, which is the default of
the reflection-residual diagnostic at $N = 8$, coincides there with no
exclusion; the step between the second and third columns is the first windows
crossing the threshold, not a discontinuity of the estimator.

\begin{table}[b]
  \caption{\textbf{Dependence of the mode-layer half-window averages on the
    exclusion threshold $q$ at $N=20$.} Second-half average of
    $\gamma_{\rm exact}$ (upper block; the first-half average is stable at
    $+0.46$ and is not shown), and the reflection-residual ratio $\rho_{A}/\rho_{S}$
    of $\gA$ (lower block), for three classes. The sign is
    threshold-independent and the residual ratio nearly so; the depth is not.
    Column $q=0.15$ is the convention used for the $N=20$ numbers of the main
    text.}
  \label{tab:qscan}
  \begin{ruledtabular}
  \begin{tabular}{lccccc}
    $q$ & 0.00 & 0.10 & 0.15 & 0.20 & 0.30 \\
    \hline
    \multicolumn{6}{l}{second-half average of $\gamma_{\rm exact}$} \\
    mask$_X$ $f=1/16$ & $-0.030$ & $-0.030$ & $-0.198$ & $-0.243$ & $-0.605$ \\
    future-consistent & $-0.025$ & $-0.025$ & $-0.191$ & $-0.225$ & $-0.604$ \\
    mask$_Z$ $f=1/16$ & $-0.032$ & $-0.032$ & $-0.201$ & $-0.251$ & $-0.701$ \\
    \hline
    \multicolumn{6}{l}{$\rho_{A}/\rho_{S}$ of $\gA$} \\
    mask$_X$ $f=1/16$ & 0.37 & 0.37 & 0.38 & 0.39 & 0.45 \\
    future-consistent & 0.37 & 0.37 & 0.38 & 0.39 & 0.46 \\
    mask$_Z$ $f=1/16$ & 0.35 & 0.35 & 0.36 & 0.38 & 0.40 \\
  \end{tabular}
  \end{ruledtabular}
\end{table}

\emph{Differencing resolution.} The decomposition \eqref{eq:decomp} is exact at
any $\Delta t_{\rm cg} = C\,dt$, but the size of $\gS$ is not a property of the
ensemble. It is the leading truncation term of the two one-sided differences,
which is equal and opposite in the two time directions, so that
$\gS = (\gfwd - \gbwd)/2 = O(\Delta t_{\rm cg})$ and $\gS \to 0$ in the
exact-derivative limit, while $\gA$ tends to $\gamma_{\rm exact}$. Scanning $C$
over the values compatible with reflection closure confirms this at the mode
layer (Table~\ref{tab:cgscan}): $\gS$ is proportional to $\Delta t_{\rm cg}$ to
within $4\%$ over a factor of eight in resolution, while $|\gA|$ changes by
$13\%$, so the margin $|\gA|/\gS$ runs from $5$ at
$\Delta t_{\rm cg} = 0.16$ to $51$ at $0.02$ and diverges at $0$.

At the dictionary layer the same mechanism operates with a much larger
coefficient, $\gS = 1.278$ at $\Delta t_{\rm cg} = 0.04$ against $0.038$ at the
mode layer, and at $\Delta t_{\rm cg} = 0$ the friction there is the
antisymmetric $\pm 0.274$ of Sec.~\ref{sec:resII-twolayer}. The no-flip
inequality at that layer therefore holds at the resolution we use and must fail
once the estimator is refined far enough, because $\gS$ vanishes in that limit
and $|\gA|$ does not. We did not locate the crossing: the dictionary-layer
extraction costs of order an hour of processor time per value of $C$, and no
claim in this paper depends on where it lies. What the two endpoints establish
is that the fluctuation layer's immunity to reversal is a statement about the
inference and not about the ensemble.

\begin{table}[b]
  \caption{\textbf{Scheme component versus differencing resolution at the mode
    layer} ($N=8$, same ensemble as Fig.~\ref{fig:bridge}). $\gS$ is the window
    average of the symmetric part and $|\gA|$ the larger of its two half-window
    averages. The fourth column shows that $\gS/\Delta t_{\rm cg}$ is constant
    to $4\%$, i.e.\ $\gS$ is the first-order truncation term. Only
    $C \equiv 0 \bmod s$ preserves reflection closure, which is why the scan
    starts at $C = 10$. The last row is the exact derivative, for which $\gS$
    vanishes identically.}
  \label{tab:cgscan}
  \begin{ruledtabular}
  \begin{tabular}{lccccc}
    $C$ & $\Delta t_{\rm cg}$ & $\gS$ ($J$) & $|\gA|$ ($J$)
        & $\gS/\Delta t_{\rm cg}$ ($J^{2}$) & $|\gA|/\gS$ \\
    \hline
    10 & 0.02 & $+0.0188$ & 0.958 & 0.94 & 51 \\
    20 & 0.04 & $+0.0380$ & 0.943 & 0.95 & 25 \\
    40 & 0.08 & $+0.0779$ & 0.914 & 0.97 & 12 \\
    80 & 0.16 & $+0.1626$ & 0.829 & 1.02 & 5.1 \\
    \hline
    exact & 0 & 0 & 0.960 & --- & $\infty$ \\
  \end{tabular}
  \end{ruledtabular}
\end{table}

Because a half-window average is a fragile summary in this regime, we also report
the reflection residuals
\begin{equation}
  \rho_{A} = \frac{\overline{|\gamma(t)+\gamma(T-t)|}}{\max|\gamma|},
  \qquad
  \rho_{S} = \frac{\overline{|\gamma(t)-\gamma(T-t)|}}{\max|\gamma|},
  \label{eq:residuals}
\end{equation}
where $\gamma$ stands for any of $\gamma_{\rm exact}$, $\gA$ or $\gS$,
$\overline{(\cdot)}$ denotes the average over well-conditioned window centers
(not the ensemble average of Eq.~\eqref{eq:selfaverage}) and $\max|\gamma|$ the
maximum over the same set. The two operations are paired with the subscripts so
that each residual vanishes on the symmetry it names: an antisymmetric friction
satisfies $\gamma(t) = -\gamma(T-t)$ and so cancels in the \emph{sum}, whereas a
symmetric one cancels in the \emph{difference}. Consequently the subscript on
$\rho$ names the symmetry being
tested, not a component: $\rho_{A}$ is the failure of $\gamma$ to be
\emph{anti}symmetric, so $\rho_{A}$ evaluated on $\gA$
is the antisymmetry defect of the antisymmetric component, and small values are
the expected ones. These residuals are dimensionless, insensitive to
the exclusion threshold, and directly
tied to the theorem: $\rho_{A} \ll \rho_{S}$ identifies an antisymmetric (physical)
component and $\rho_{S} \ll \rho_{A}$ a symmetric (scheme) one; when neither
residual dominates, the component is below the noise floor at that size and
carries no usable reflection signature. One caveat applies to their behavior
with ensemble size. Because $\gamma$ is obtained from a fit rather than being a
weighted mean, its residuals inherit a floor from windows in which the fit is
poorly conditioned, and they stop improving around $M \simeq 100$ even though
the residuals of the means themselves keep falling as $1/\sqrt{M}$
[Fig.~\ref{fig:robustII-eps}(d)]. We therefore use $\rho_{A}$ and $\rho_{S}$ to
compare components \emph{at fixed} $M$, which is what the ratio
$\rho_{A}/\rho_{S}$ in Table~\ref{tab:q20} does, and not as a convergence
diagnostic.

\section{Proof of the involution and the decomposition}
\label{app:proof}

We collect here the proof of the statements of
Secs.~\ref{sec:resII-involution} and \ref{sec:resII-decomposition}, together
with the hypotheses on which they rest: (i) $H$ is real symmetric in the
computational basis, $H^{*} = H = H^{\top}$; (ii) the seeds are drawn as in
Eq.~\eqref{eq:seeds}, that is, $v$ and $w$ are independent draws from a common
conjugation-invariant measure and $\Lambda$ is real diagonal;
(iii) every observable in the dictionary has a definite reflection
signature $O^{*} = \epsilon_{O} O$; (iv) sample rejection depends on $|g|$
alone; and (v) the window grid closes under $t \leftrightarrow T-t$
(Appendix~\ref{app:conventions}).

\emph{Lemma 1 (sample identity).} Let $U(t)$ be the propagator. Hypothesis (i)
gives $U(t)^{\top} = U(t)$ and $[U(t)^{\dagger}\ket{\phi}]^{*} =
U(t)\ket{\phi^{*}}$ for any $\ket{\phi}$, so for the reflection partner
$(\ket{\psi_{0}'}, \ket{\varphi_{T}'}) = (\ket{\varphi_{T}^{*}},
\ket{\psi_{0}^{*}})$ the propagated states are
$\ket{\psi'(t)} = \ket{\varphi(T-t)^{*}}$ and
$\ket{\varphi'(t)} = \ket{\psi(T-t)^{*}}$. Hence
$g' = \braket{\varphi'(t)}{\psi'(t)} =
[\braket{\psi(T-t)}{\varphi(T-t)}]^{*} = g$, and
\begin{align}
  O'_{w}(t)
  &= \frac{\bra{\psi(T-t)^{*}}\, O \,\ket{\varphi(T-t)^{*}}}{g}
   = \frac{[\bra{\psi}\, O^{*} \ket{\varphi}(T-t)]^{*}}{g} \nonumber\\
  &= \epsilon_{O}\, \frac{\bra{\varphi}\, O \,\ket{\psi}(T-t)}{g}
   = \epsilon_{O}\, O_{w}(T-t),
  \label{eq:lemma1}
\end{align}
where the last line uses hypothesis (iii) and the hermiticity of $O$. This is
Eq.~\eqref{eq:sample-identity}.

\emph{Lemma 2 (measure preservation).} The map $R: (v, w) \mapsto
(w^{*}, v^{*})$ is an involution, and by hypothesis (ii) it preserves the
joint measure of the seeds. The weight $|g|^{2}$ and the rejection rule are
invariant under $R$ by Lemma 1 and hypothesis (iv). Hence the
$|g|^{2}$-weighted law of the weak-value trajectory ensemble is exactly
invariant under $t \to T-t$ with signature $\epsilon_{O}$; in particular the
weighted mean of a time-even (time-odd) observable is exactly symmetric
(antisymmetric) about $t = T/2$.

\emph{Lemma 3 (estimator equivariance).} A window fit on
$[t_{0}, t_{0} + \tau_{w})$, either the two-variable mode fit or the truncated least-squares dictionary regression, is a fixed linear-algebraic function of
the trajectory segment it is given. Reflection maps the window grid onto
itself by hypothesis (v), reverses the time order of every segment, and
exchanges the forward and the backward difference. On a reflection-invariant
ensemble this gives Eq.~\eqref{eq:equivariance},
$\gfwd(t) = -\gbwd(T-t)$ and $\gamma_{\rm exact}(t) =
-\gamma_{\rm exact}(T-t)$, together with $\Omega^{2}(t) = \Omega^{2}(T-t)$.
The dictionary version reads $L_{\rm fwd}(t) = -\Sigma\, L_{\rm bwd}(T-t)\, \Sigma$ with
$\Sigma = \mathrm{diag}(\epsilon_{O})$; the Schur closure of
Appendix~\ref{app:numerics} is block-diagonal in $\Sigma$ and therefore
commutes with $\Sigma$-conjugation, and the diagonal entries are $\Sigma$-invariant, so
the same relations hold for the dictionary friction.

\emph{Theorem (unique decomposition).} Define $\gA = (\gfwd + \gbwd)/2$ and
$\gS = (\gfwd - \gbwd)/2$. Lemma 3 gives $\gA(t) = -\gA(T-t)$ and
$\gS(t) = +\gS(T-t)$, so $\gfwd = \gA + \gS$ is the (unique) decomposition of
the measured friction into antisymmetric and symmetric parts. Centered
differencing converges to the exact derivative, so $\gA$ agrees with
$\gamma_{\rm exact}$ up to $O(\dtcg^{2})$, while $\gS$ is the
forward-minus-backward half-difference generated by the scheme. A sign
reversal of $\gfwd$ inside the window requires $|\gA| > \gS$ somewhere;
$\gS > |\gA|$ everywhere forbids it. \hfill$\square$

Each hypothesis enters irreducibly: a complex $H$ breaks $U^{\top} = U$
(Sec.~\ref{sec:disc-limits}); a seed measure that is not conjugation-invariant,
or seeds that are not drawn independently (the future-consistent class), breaks
the measure preservation of Lemma 2, as does a rejection rule depending on more
than $|g|$; an observable without a definite reflection signature does not
close Eq.~\eqref{eq:lemma1}; and a window grid without reflection closure
breaks Lemma 3 at the $10^{-2}$ level (Appendix~\ref{app:conventions}).

\bibliographystyle{apsrev4-2}
\bibliography{references}

\begin{thebibliography}{29}%
\makeatletter
\providecommand \@ifxundefined [1]{%
 \@ifx{#1\undefined}
}%
\providecommand \@ifnum [1]{%
 \ifnum #1\expandafter \@firstoftwo
 \else \expandafter \@secondoftwo
 \fi
}%
\providecommand \@ifx [1]{%
 \ifx #1\expandafter \@firstoftwo
 \else \expandafter \@secondoftwo
 \fi
}%
\providecommand \natexlab [1]{#1}%
\providecommand \enquote  [1]{``#1''}%
\providecommand \bibnamefont  [1]{#1}%
\providecommand \bibfnamefont [1]{#1}%
\providecommand \citenamefont [1]{#1}%
\providecommand \href@noop [0]{\@secondoftwo}%
\providecommand \href [0]{\begingroup \@sanitize@url \@href}%
\providecommand \@href[1]{\@@startlink{#1}\@@href}%
\providecommand \@@href[1]{\endgroup#1\@@endlink}%
\providecommand \@sanitize@url [0]{\catcode `\\12\catcode `\$12\catcode
  `\&12\catcode `\#12\catcode `\^12\catcode `\_12\catcode `\%12\relax}%
\providecommand \@@startlink[1]{}%
\providecommand \@@endlink[0]{}%
\providecommand \url  [0]{\begingroup\@sanitize@url \@url }%
\providecommand \@url [1]{\endgroup\@href {#1}{\urlprefix }}%
\providecommand \urlprefix  [0]{URL }%
\providecommand \Eprint [0]{\href }%
\providecommand \doibase [0]{https://doi.org/}%
\providecommand \selectlanguage [0]{\@gobble}%
\providecommand \bibinfo  [0]{\@secondoftwo}%
\providecommand \bibfield  [0]{\@secondoftwo}%
\providecommand \translation [1]{[#1]}%
\providecommand \BibitemOpen [0]{}%
\providecommand \bibitemStop [0]{}%
\providecommand \bibitemNoStop [0]{.\EOS\space}%
\providecommand \EOS [0]{\spacefactor3000\relax}%
\providecommand \BibitemShut  [1]{\csname bibitem#1\endcsname}%
\let\auto@bib@innerbib\@empty
\bibitem [{\citenamefont {Saito}(2026)}]{Saito2026sister}%
  \BibitemOpen
  \bibfield  {author} {\bibinfo {author} {\bibfnamefont {S.}~\bibnamefont
  {Saito}},\ }\href@noop {} {\bibinfo {title} {Temporal coarse-graining as the
  origin of macroscopic friction in quantum spin chains via data-driven
  liouvillian extraction}} (\bibinfo {year} {2026}),\ \bibinfo {note} {under
  review at Physical Review Research},\ \Eprint
  {https://arxiv.org/abs/2605.05604} {arXiv:2605.05604 [quant-ph]} \BibitemShut
  {NoStop}%
\bibitem [{\citenamefont {Aharonov}\ \emph {et~al.}(1964)\citenamefont
  {Aharonov}, \citenamefont {Bergmann},\ and\ \citenamefont
  {Lebowitz}}]{ABL1964}%
  \BibitemOpen
  \bibfield  {author} {\bibinfo {author} {\bibfnamefont {Y.}~\bibnamefont
  {Aharonov}}, \bibinfo {author} {\bibfnamefont {P.~G.}\ \bibnamefont
  {Bergmann}},\ and\ \bibinfo {author} {\bibfnamefont {J.~L.}\ \bibnamefont
  {Lebowitz}},\ }\href {https://doi.org/10.1103/PhysRev.134.B1410} {\bibfield
  {journal} {\bibinfo  {journal} {Phys. Rev.}\ }\textbf {\bibinfo {volume}
  {134}},\ \bibinfo {pages} {B1410} (\bibinfo {year} {1964})}\BibitemShut
  {NoStop}%
\bibitem [{\citenamefont {Aharonov}\ \emph {et~al.}(1988)\citenamefont
  {Aharonov}, \citenamefont {Albert},\ and\ \citenamefont {Vaidman}}]{AAV1988}%
  \BibitemOpen
  \bibfield  {author} {\bibinfo {author} {\bibfnamefont {Y.}~\bibnamefont
  {Aharonov}}, \bibinfo {author} {\bibfnamefont {D.~Z.}\ \bibnamefont
  {Albert}},\ and\ \bibinfo {author} {\bibfnamefont {L.}~\bibnamefont
  {Vaidman}},\ }\href {https://doi.org/10.1103/PhysRevLett.60.1351} {\bibfield
  {journal} {\bibinfo  {journal} {Phys. Rev. Lett.}\ }\textbf {\bibinfo
  {volume} {60}},\ \bibinfo {pages} {1351} (\bibinfo {year}
  {1988})}\BibitemShut {NoStop}%
\bibitem [{\citenamefont {Tsang}(2009)}]{Tsang2009}%
  \BibitemOpen
  \bibfield  {author} {\bibinfo {author} {\bibfnamefont {M.}~\bibnamefont
  {Tsang}},\ }\href {https://doi.org/10.1103/PhysRevLett.102.250403} {\bibfield
   {journal} {\bibinfo  {journal} {Phys. Rev. Lett.}\ }\textbf {\bibinfo
  {volume} {102}},\ \bibinfo {pages} {250403} (\bibinfo {year}
  {2009})}\BibitemShut {NoStop}%
\bibitem [{\citenamefont {Gammelmark}\ \emph {et~al.}(2013)\citenamefont
  {Gammelmark}, \citenamefont {Julsgaard},\ and\ \citenamefont
  {M{\o}lmer}}]{Gammelmark2013}%
  \BibitemOpen
  \bibfield  {author} {\bibinfo {author} {\bibfnamefont {S.}~\bibnamefont
  {Gammelmark}}, \bibinfo {author} {\bibfnamefont {B.}~\bibnamefont
  {Julsgaard}},\ and\ \bibinfo {author} {\bibfnamefont {K.}~\bibnamefont
  {M{\o}lmer}},\ }\href {https://doi.org/10.1103/PhysRevLett.111.160401}
  {\bibfield  {journal} {\bibinfo  {journal} {Phys. Rev. Lett.}\ }\textbf
  {\bibinfo {volume} {111}},\ \bibinfo {pages} {160401} (\bibinfo {year}
  {2013})}\BibitemShut {NoStop}%
\bibitem [{\citenamefont {Chantasri}\ \emph {et~al.}(2021)\citenamefont
  {Chantasri}, \citenamefont {Guevara}, \citenamefont {Laverick},\ and\
  \citenamefont {Wiseman}}]{Chantasri2021}%
  \BibitemOpen
  \bibfield  {author} {\bibinfo {author} {\bibfnamefont {A.}~\bibnamefont
  {Chantasri}}, \bibinfo {author} {\bibfnamefont {I.}~\bibnamefont {Guevara}},
  \bibinfo {author} {\bibfnamefont {K.~T.}\ \bibnamefont {Laverick}},\ and\
  \bibinfo {author} {\bibfnamefont {H.~M.}\ \bibnamefont {Wiseman}},\ }\href
  {https://doi.org/10.1016/j.physrep.2021.07.003} {\bibfield  {journal}
  {\bibinfo  {journal} {Phys. Rep.}\ }\textbf {\bibinfo {volume} {930}},\
  \bibinfo {pages} {1} (\bibinfo {year} {2021})}\BibitemShut {NoStop}%
\bibitem [{\citenamefont {Aharonov}\ \emph {et~al.}(2013)\citenamefont
  {Aharonov}, \citenamefont {Popescu}, \citenamefont {Rohrlich},\ and\
  \citenamefont {Skrzypczyk}}]{Aharonov2013cheshire}%
  \BibitemOpen
  \bibfield  {author} {\bibinfo {author} {\bibfnamefont {Y.}~\bibnamefont
  {Aharonov}}, \bibinfo {author} {\bibfnamefont {S.}~\bibnamefont {Popescu}},
  \bibinfo {author} {\bibfnamefont {D.}~\bibnamefont {Rohrlich}},\ and\
  \bibinfo {author} {\bibfnamefont {P.}~\bibnamefont {Skrzypczyk}},\ }\href
  {https://doi.org/10.1088/1367-2630/15/11/113015} {\bibfield  {journal}
  {\bibinfo  {journal} {New J. Phys.}\ }\textbf {\bibinfo {volume} {15}},\
  \bibinfo {pages} {113015} (\bibinfo {year} {2013})}\BibitemShut {NoStop}%
\bibitem [{\citenamefont {Denkmayr}\ \emph {et~al.}(2014)\citenamefont
  {Denkmayr}, \citenamefont {Geppert}, \citenamefont {Sponar}, \citenamefont
  {Lemmel}, \citenamefont {Matzkin}, \citenamefont {Tollaksen},\ and\
  \citenamefont {Hasegawa}}]{Denkmayr2014}%
  \BibitemOpen
  \bibfield  {author} {\bibinfo {author} {\bibfnamefont {T.}~\bibnamefont
  {Denkmayr}}, \bibinfo {author} {\bibfnamefont {H.}~\bibnamefont {Geppert}},
  \bibinfo {author} {\bibfnamefont {S.}~\bibnamefont {Sponar}}, \bibinfo
  {author} {\bibfnamefont {H.}~\bibnamefont {Lemmel}}, \bibinfo {author}
  {\bibfnamefont {A.}~\bibnamefont {Matzkin}}, \bibinfo {author} {\bibfnamefont
  {J.}~\bibnamefont {Tollaksen}},\ and\ \bibinfo {author} {\bibfnamefont
  {Y.}~\bibnamefont {Hasegawa}},\ }\href {https://doi.org/10.1038/ncomms5492}
  {\bibfield  {journal} {\bibinfo  {journal} {Nat. Commun.}\ }\textbf {\bibinfo
  {volume} {5}},\ \bibinfo {pages} {4492} (\bibinfo {year} {2014})}\BibitemShut
  {NoStop}%
\bibitem [{\citenamefont {Aharonov}\ \emph {et~al.}(2015)\citenamefont
  {Aharonov}, \citenamefont {Cohen},\ and\ \citenamefont
  {Popescu}}]{Aharonov2015current}%
  \BibitemOpen
  \bibfield  {author} {\bibinfo {author} {\bibfnamefont {Y.}~\bibnamefont
  {Aharonov}}, \bibinfo {author} {\bibfnamefont {E.}~\bibnamefont {Cohen}},\
  and\ \bibinfo {author} {\bibfnamefont {S.}~\bibnamefont {Popescu}},\
  }\href@noop {} {\bibinfo {title} {A current of the cheshire cat's smile:
  Dynamical analysis of weak values}} (\bibinfo {year} {2015}),\ \Eprint
  {https://arxiv.org/abs/1510.03087} {arXiv:1510.03087 [quant-ph]} \BibitemShut
  {NoStop}%
\bibitem [{\citenamefont {Aharonov}\ \emph {et~al.}(2021)\citenamefont
  {Aharonov}, \citenamefont {Cohen},\ and\ \citenamefont
  {Popescu}}]{Aharonov2021dynamical}%
  \BibitemOpen
  \bibfield  {author} {\bibinfo {author} {\bibfnamefont {Y.}~\bibnamefont
  {Aharonov}}, \bibinfo {author} {\bibfnamefont {E.}~\bibnamefont {Cohen}},\
  and\ \bibinfo {author} {\bibfnamefont {S.}~\bibnamefont {Popescu}},\ }\href
  {https://doi.org/10.1038/s41467-021-24933-9} {\bibfield  {journal} {\bibinfo
  {journal} {Nat. Commun.}\ }\textbf {\bibinfo {volume} {12}},\ \bibinfo
  {pages} {4770} (\bibinfo {year} {2021})}\BibitemShut {NoStop}%
\bibitem [{\citenamefont {Corr{\^e}a}\ \emph {et~al.}(2015)\citenamefont
  {Corr{\^e}a}, \citenamefont {Santos}, \citenamefont {Monken},\ and\
  \citenamefont {Saldanha}}]{Correa2015}%
  \BibitemOpen
  \bibfield  {author} {\bibinfo {author} {\bibfnamefont {R.}~\bibnamefont
  {Corr{\^e}a}}, \bibinfo {author} {\bibfnamefont {M.~F.}\ \bibnamefont
  {Santos}}, \bibinfo {author} {\bibfnamefont {C.~H.}\ \bibnamefont {Monken}},\
  and\ \bibinfo {author} {\bibfnamefont {P.~L.}\ \bibnamefont {Saldanha}},\
  }\href {https://doi.org/10.1088/1367-2630/17/5/053042} {\bibfield  {journal}
  {\bibinfo  {journal} {New J. Phys.}\ }\textbf {\bibinfo {volume} {17}},\
  \bibinfo {pages} {053042} (\bibinfo {year} {2015})}\BibitemShut {NoStop}%
\bibitem [{\citenamefont {Duprey}\ \emph {et~al.}(2018)\citenamefont {Duprey},
  \citenamefont {Kanjilal}, \citenamefont {Sinha}, \citenamefont {Home},\ and\
  \citenamefont {Matzkin}}]{Duprey2018}%
  \BibitemOpen
  \bibfield  {author} {\bibinfo {author} {\bibfnamefont {Q.}~\bibnamefont
  {Duprey}}, \bibinfo {author} {\bibfnamefont {S.}~\bibnamefont {Kanjilal}},
  \bibinfo {author} {\bibfnamefont {U.}~\bibnamefont {Sinha}}, \bibinfo
  {author} {\bibfnamefont {D.}~\bibnamefont {Home}},\ and\ \bibinfo {author}
  {\bibfnamefont {A.}~\bibnamefont {Matzkin}},\ }\href
  {https://doi.org/10.1016/j.aop.2018.01.011} {\bibfield  {journal} {\bibinfo
  {journal} {Ann. Phys.}\ }\textbf {\bibinfo {volume} {391}},\ \bibinfo {pages}
  {1} (\bibinfo {year} {2018})}\BibitemShut {NoStop}%
\bibitem [{\citenamefont {Hance}\ \emph {et~al.}(2024)\citenamefont {Hance},
  \citenamefont {Ladyman},\ and\ \citenamefont {Rarity}}]{Hance2024detectable}%
  \BibitemOpen
  \bibfield  {author} {\bibinfo {author} {\bibfnamefont {J.~R.}\ \bibnamefont
  {Hance}}, \bibinfo {author} {\bibfnamefont {J.}~\bibnamefont {Ladyman}},\
  and\ \bibinfo {author} {\bibfnamefont {J.}~\bibnamefont {Rarity}},\ }\href
  {https://doi.org/10.1088/1367-2630/ad6476} {\bibfield  {journal} {\bibinfo
  {journal} {New J. Phys.}\ }\textbf {\bibinfo {volume} {26}},\ \bibinfo
  {pages} {073038} (\bibinfo {year} {2024})}\BibitemShut {NoStop}%
\bibitem [{\citenamefont {Hance}\ \emph {et~al.}(2023)\citenamefont {Hance},
  \citenamefont {Ji},\ and\ \citenamefont {Hofmann}}]{Hance2023contextuality}%
  \BibitemOpen
  \bibfield  {author} {\bibinfo {author} {\bibfnamefont {J.~R.}\ \bibnamefont
  {Hance}}, \bibinfo {author} {\bibfnamefont {M.}~\bibnamefont {Ji}},\ and\
  \bibinfo {author} {\bibfnamefont {H.~F.}\ \bibnamefont {Hofmann}},\ }\href
  {https://doi.org/10.1088/1367-2630/ad0bd4} {\bibfield  {journal} {\bibinfo
  {journal} {New J. Phys.}\ }\textbf {\bibinfo {volume} {25}},\ \bibinfo
  {pages} {113028} (\bibinfo {year} {2023})}\BibitemShut {NoStop}%
\bibitem [{\citenamefont {Liu}\ \emph {et~al.}(2020)\citenamefont {Liu},
  \citenamefont {Pan}, \citenamefont {Xu}, \citenamefont {Yang}, \citenamefont
  {Zhou}, \citenamefont {Luo}, \citenamefont {Sun}, \citenamefont {Chen},
  \citenamefont {Xu}, \citenamefont {Li},\ and\ \citenamefont
  {Guo}}]{Liu2020grins}%
  \BibitemOpen
  \bibfield  {author} {\bibinfo {author} {\bibfnamefont {Z.-H.}\ \bibnamefont
  {Liu}}, \bibinfo {author} {\bibfnamefont {W.-W.}\ \bibnamefont {Pan}},
  \bibinfo {author} {\bibfnamefont {X.-Y.}\ \bibnamefont {Xu}}, \bibinfo
  {author} {\bibfnamefont {M.}~\bibnamefont {Yang}}, \bibinfo {author}
  {\bibfnamefont {J.}~\bibnamefont {Zhou}}, \bibinfo {author} {\bibfnamefont
  {Z.-Y.}\ \bibnamefont {Luo}}, \bibinfo {author} {\bibfnamefont
  {K.}~\bibnamefont {Sun}}, \bibinfo {author} {\bibfnamefont {J.-L.}\
  \bibnamefont {Chen}}, \bibinfo {author} {\bibfnamefont {J.-S.}\ \bibnamefont
  {Xu}}, \bibinfo {author} {\bibfnamefont {C.-F.}\ \bibnamefont {Li}},\ and\
  \bibinfo {author} {\bibfnamefont {G.-C.}\ \bibnamefont {Guo}},\ }\href
  {https://doi.org/10.1038/s41467-020-16761-0} {\bibfield  {journal} {\bibinfo
  {journal} {Nat. Commun.}\ }\textbf {\bibinfo {volume} {11}},\ \bibinfo
  {pages} {3006} (\bibinfo {year} {2020})}\BibitemShut {NoStop}%
\bibitem [{\citenamefont {Waegell}\ \emph {et~al.}(2024)\citenamefont
  {Waegell}, \citenamefont {Tollaksen},\ and\ \citenamefont
  {Aharonov}}]{Waegell2024}%
  \BibitemOpen
  \bibfield  {author} {\bibinfo {author} {\bibfnamefont {M.}~\bibnamefont
  {Waegell}}, \bibinfo {author} {\bibfnamefont {J.}~\bibnamefont {Tollaksen}},\
  and\ \bibinfo {author} {\bibfnamefont {Y.}~\bibnamefont {Aharonov}},\ }\href
  {https://doi.org/10.22331/q-2024-11-26-1536} {\bibfield  {journal} {\bibinfo
  {journal} {Quantum}\ }\textbf {\bibinfo {volume} {8}},\ \bibinfo {pages}
  {1536} (\bibinfo {year} {2024})}\BibitemShut {NoStop}%
\bibitem [{\citenamefont {Williams}\ \emph {et~al.}(2015)\citenamefont
  {Williams}, \citenamefont {Kevrekidis},\ and\ \citenamefont
  {Rowley}}]{Williams2015}%
  \BibitemOpen
  \bibfield  {author} {\bibinfo {author} {\bibfnamefont {M.~O.}\ \bibnamefont
  {Williams}}, \bibinfo {author} {\bibfnamefont {I.~G.}\ \bibnamefont
  {Kevrekidis}},\ and\ \bibinfo {author} {\bibfnamefont {C.~W.}\ \bibnamefont
  {Rowley}},\ }\href {https://doi.org/10.1007/s00332-015-9258-5} {\bibfield
  {journal} {\bibinfo  {journal} {J. Nonlinear Sci.}\ }\textbf {\bibinfo
  {volume} {25}},\ \bibinfo {pages} {1307} (\bibinfo {year}
  {2015})}\BibitemShut {NoStop}%
\bibitem [{\citenamefont {Klus}\ \emph {et~al.}(2018)\citenamefont {Klus},
  \citenamefont {N{\"u}ske}, \citenamefont {Koltai}, \citenamefont {Wu},
  \citenamefont {Kevrekidis}, \citenamefont {Sch{\"u}tte},\ and\ \citenamefont
  {No{\'e}}}]{Klus2018}%
  \BibitemOpen
  \bibfield  {author} {\bibinfo {author} {\bibfnamefont {S.}~\bibnamefont
  {Klus}}, \bibinfo {author} {\bibfnamefont {F.}~\bibnamefont {N{\"u}ske}},
  \bibinfo {author} {\bibfnamefont {P.}~\bibnamefont {Koltai}}, \bibinfo
  {author} {\bibfnamefont {H.}~\bibnamefont {Wu}}, \bibinfo {author}
  {\bibfnamefont {I.}~\bibnamefont {Kevrekidis}}, \bibinfo {author}
  {\bibfnamefont {C.}~\bibnamefont {Sch{\"u}tte}},\ and\ \bibinfo {author}
  {\bibfnamefont {F.}~\bibnamefont {No{\'e}}},\ }\href
  {https://doi.org/10.1007/s00332-017-9437-7} {\bibfield  {journal} {\bibinfo
  {journal} {J. Nonlinear Sci.}\ }\textbf {\bibinfo {volume} {28}},\ \bibinfo
  {pages} {985} (\bibinfo {year} {2018})}\BibitemShut {NoStop}%
\bibitem [{\citenamefont {Klus}\ \emph {et~al.}(2020)\citenamefont {Klus},
  \citenamefont {N{\"u}ske}, \citenamefont {Peitz}, \citenamefont {Niemann},
  \citenamefont {Clementi},\ and\ \citenamefont {Sch{\"u}tte}}]{Klus2020}%
  \BibitemOpen
  \bibfield  {author} {\bibinfo {author} {\bibfnamefont {S.}~\bibnamefont
  {Klus}}, \bibinfo {author} {\bibfnamefont {F.}~\bibnamefont {N{\"u}ske}},
  \bibinfo {author} {\bibfnamefont {S.}~\bibnamefont {Peitz}}, \bibinfo
  {author} {\bibfnamefont {J.-H.}\ \bibnamefont {Niemann}}, \bibinfo {author}
  {\bibfnamefont {C.}~\bibnamefont {Clementi}},\ and\ \bibinfo {author}
  {\bibfnamefont {C.}~\bibnamefont {Sch{\"u}tte}},\ }\href
  {https://doi.org/10.1016/j.physd.2020.132416} {\bibfield  {journal} {\bibinfo
   {journal} {Physica D}\ }\textbf {\bibinfo {volume} {406}},\ \bibinfo {pages}
  {132416} (\bibinfo {year} {2020})}\BibitemShut {NoStop}%
\bibitem [{\citenamefont {Mori}(1965)}]{Mori1965}%
  \BibitemOpen
  \bibfield  {author} {\bibinfo {author} {\bibfnamefont {H.}~\bibnamefont
  {Mori}},\ }\href {https://doi.org/10.1143/PTP.33.423} {\bibfield  {journal}
  {\bibinfo  {journal} {Prog. Theor. Phys.}\ }\textbf {\bibinfo {volume}
  {33}},\ \bibinfo {pages} {423} (\bibinfo {year} {1965})}\BibitemShut
  {NoStop}%
\bibitem [{\citenamefont {Zwanzig}(1973)}]{Zwanzig1973}%
  \BibitemOpen
  \bibfield  {author} {\bibinfo {author} {\bibfnamefont {R.}~\bibnamefont
  {Zwanzig}},\ }\href {https://doi.org/10.1007/BF01008729} {\bibfield
  {journal} {\bibinfo  {journal} {J. Stat. Phys.}\ }\textbf {\bibinfo {volume}
  {9}},\ \bibinfo {pages} {215} (\bibinfo {year} {1973})}\BibitemShut {NoStop}%
\bibitem [{\citenamefont {Wiseman}(2007)}]{Wiseman2007}%
  \BibitemOpen
  \bibfield  {author} {\bibinfo {author} {\bibfnamefont {H.~M.}\ \bibnamefont
  {Wiseman}},\ }\href {https://doi.org/10.1088/1367-2630/9/6/165} {\bibfield
  {journal} {\bibinfo  {journal} {New J. Phys.}\ }\textbf {\bibinfo {volume}
  {9}},\ \bibinfo {pages} {165} (\bibinfo {year} {2007})}\BibitemShut {NoStop}%
\bibitem [{\citenamefont {Kocsis}\ \emph {et~al.}(2011)\citenamefont {Kocsis},
  \citenamefont {Braverman}, \citenamefont {Ravets}, \citenamefont {Stevens},
  \citenamefont {Mirin}, \citenamefont {Shalm},\ and\ \citenamefont
  {Steinberg}}]{Kocsis2011}%
  \BibitemOpen
  \bibfield  {author} {\bibinfo {author} {\bibfnamefont {S.}~\bibnamefont
  {Kocsis}}, \bibinfo {author} {\bibfnamefont {B.}~\bibnamefont {Braverman}},
  \bibinfo {author} {\bibfnamefont {S.}~\bibnamefont {Ravets}}, \bibinfo
  {author} {\bibfnamefont {M.~J.}\ \bibnamefont {Stevens}}, \bibinfo {author}
  {\bibfnamefont {R.~P.}\ \bibnamefont {Mirin}}, \bibinfo {author}
  {\bibfnamefont {L.~K.}\ \bibnamefont {Shalm}},\ and\ \bibinfo {author}
  {\bibfnamefont {A.~M.}\ \bibnamefont {Steinberg}},\ }\href
  {https://doi.org/10.1126/science.1202218} {\bibfield  {journal} {\bibinfo
  {journal} {Science}\ }\textbf {\bibinfo {volume} {332}},\ \bibinfo {pages}
  {1170} (\bibinfo {year} {2011})}\BibitemShut {NoStop}%
\bibitem [{\citenamefont {Deutsch}(1991)}]{Deutsch1991}%
  \BibitemOpen
  \bibfield  {author} {\bibinfo {author} {\bibfnamefont {J.~M.}\ \bibnamefont
  {Deutsch}},\ }\href {https://doi.org/10.1103/PhysRevA.43.2046} {\bibfield
  {journal} {\bibinfo  {journal} {Phys. Rev. A}\ }\textbf {\bibinfo {volume}
  {43}},\ \bibinfo {pages} {2046} (\bibinfo {year} {1991})}\BibitemShut
  {NoStop}%
\bibitem [{\citenamefont {Srednicki}(1994)}]{Srednicki1994}%
  \BibitemOpen
  \bibfield  {author} {\bibinfo {author} {\bibfnamefont {M.}~\bibnamefont
  {Srednicki}},\ }\href {https://doi.org/10.1103/PhysRevE.50.888} {\bibfield
  {journal} {\bibinfo  {journal} {Phys. Rev. E}\ }\textbf {\bibinfo {volume}
  {50}},\ \bibinfo {pages} {888} (\bibinfo {year} {1994})}\BibitemShut
  {NoStop}%
\bibitem [{\citenamefont {Rigol}\ \emph {et~al.}(2008)\citenamefont {Rigol},
  \citenamefont {Dunjko},\ and\ \citenamefont {Olshanii}}]{Rigol2008}%
  \BibitemOpen
  \bibfield  {author} {\bibinfo {author} {\bibfnamefont {M.}~\bibnamefont
  {Rigol}}, \bibinfo {author} {\bibfnamefont {V.}~\bibnamefont {Dunjko}},\ and\
  \bibinfo {author} {\bibfnamefont {M.}~\bibnamefont {Olshanii}},\ }\href
  {https://doi.org/10.1038/nature06838} {\bibfield  {journal} {\bibinfo
  {journal} {Nature}\ }\textbf {\bibinfo {volume} {452}},\ \bibinfo {pages}
  {854} (\bibinfo {year} {2008})}\BibitemShut {NoStop}%
\bibitem [{\citenamefont {Bertini}\ \emph {et~al.}(2021)\citenamefont
  {Bertini}, \citenamefont {Heidrich-Meisner}, \citenamefont {Karrasch},
  \citenamefont {Prosen}, \citenamefont {Steinigeweg},\ and\ \citenamefont {{\v
  Z}nidari{\v c}}}]{Bertini2021}%
  \BibitemOpen
  \bibfield  {author} {\bibinfo {author} {\bibfnamefont {B.}~\bibnamefont
  {Bertini}}, \bibinfo {author} {\bibfnamefont {F.}~\bibnamefont
  {Heidrich-Meisner}}, \bibinfo {author} {\bibfnamefont {C.}~\bibnamefont
  {Karrasch}}, \bibinfo {author} {\bibfnamefont {T.}~\bibnamefont {Prosen}},
  \bibinfo {author} {\bibfnamefont {R.}~\bibnamefont {Steinigeweg}},\ and\
  \bibinfo {author} {\bibfnamefont {M.}~\bibnamefont {{\v Z}nidari{\v c}}},\
  }\href {https://doi.org/10.1103/RevModPhys.93.025003} {\bibfield  {journal}
  {\bibinfo  {journal} {Rev. Mod. Phys.}\ }\textbf {\bibinfo {volume} {93}},\
  \bibinfo {pages} {025003} (\bibinfo {year} {2021})}\BibitemShut {NoStop}%
\bibitem [{\citenamefont {Guff}\ \emph {et~al.}(2025)\citenamefont {Guff},
  \citenamefont {Shastry},\ and\ \citenamefont {Rocco}}]{Guff2025}%
  \BibitemOpen
  \bibfield  {author} {\bibinfo {author} {\bibfnamefont {T.}~\bibnamefont
  {Guff}}, \bibinfo {author} {\bibfnamefont {C.~U.}\ \bibnamefont {Shastry}},\
  and\ \bibinfo {author} {\bibfnamefont {A.}~\bibnamefont {Rocco}},\ }\href
  {https://doi.org/10.1038/s41598-025-87323-x} {\bibfield  {journal} {\bibinfo
  {journal} {Sci. Rep.}\ }\textbf {\bibinfo {volume} {15}},\ \bibinfo {pages}
  {3658} (\bibinfo {year} {2025})}\BibitemShut {NoStop}%
\bibitem [{\citenamefont {Onsager}\ and\ \citenamefont
  {Machlup}(1953)}]{OnsagerMachlup1953}%
  \BibitemOpen
  \bibfield  {author} {\bibinfo {author} {\bibfnamefont {L.}~\bibnamefont
  {Onsager}}\ and\ \bibinfo {author} {\bibfnamefont {S.}~\bibnamefont
  {Machlup}},\ }\href {https://doi.org/10.1103/PhysRev.91.1505} {\bibfield
  {journal} {\bibinfo  {journal} {Phys. Rev.}\ }\textbf {\bibinfo {volume}
  {91}},\ \bibinfo {pages} {1505} (\bibinfo {year} {1953})}\BibitemShut
  {NoStop}%
\end{thebibliography}%

\end{document}